\documentclass[twocolumn,superscriptaddress,showpacs,nofootinbib,notitlepage]{aastex631}
\usepackage{graphicx,amsmath,amsfonts,amssymb,multirow,mathrsfs,ulem}
\usepackage{hyperref}
\usepackage{epstopdf}
\begin{document}

\title{Constraining and comparing the dynamical dark energy and $\boldsymbol{f}(\boldsymbol{R})$ modified gravity models \\with cosmological distance measurements} 

\correspondingauthor{Yan Gong}
\email{Email: gongyan@bao.ac.cn}

\author[0009-0006-0803-0505]{Shuai Feng}
\affiliation{National Astronomical Observatories, Chinese Academy of Sciences, Beijing 100101, People's Republic of China}
\affiliation{University of Chinese Academy of Sciences, Beijing 100049, People's Republic of China}

\author[0000-0003-0709-0101]{Yan Gong}
\affiliation{National Astronomical Observatories, Chinese Academy of Sciences, Beijing 100101, People's Republic of China}
\affiliation{University of Chinese Academy of Sciences, Beijing 100049, People's Republic of China}
\affiliation{Science Center for CSST, National Astronomical Observatories, CAS, 20A Datun Road, Beijing 100101, China}

\author[0000-0002-2552-7277]{Xiaohui Liu}
\affiliation{National Astronomical Observatories, Chinese Academy of Sciences, Beijing 100101, People's Republic of China}
\affiliation{University of Chinese Academy of Sciences, Beijing 100049, People's Republic of China}

\author[0009-0001-7455-6947]{Jun-Hui Yan}
\affiliation{National Astronomical Observatories, Chinese Academy of Sciences, Beijing 100101, People's Republic of China}
\affiliation{University of Chinese Academy of Sciences, Beijing 100049, People's Republic of China}

\author[0000-0001-6475-8863]{Xuelei Chen}
\affiliation{National Astronomical Observatories, Chinese Academy of Sciences, Beijing 100101, People's Republic of China}
\affiliation{University of Chinese Academy of Sciences, Beijing 100049, People's Republic of China}
\affiliation{Department of Physics, College of Sciences, Northeastern University, Shenyang 110819, China}
\affiliation{Centre for High Energy Physics, Peking University, Beijing 100871, People's Republic of China}
\affiliation{State Key Laboratory of Radio Astronomy and Technology, China}

\begin{abstract}
We constrain and compare the $w_{0}w_{a}$CDM dynamical dark energy model and three $f(R)$ modified gravity models using  the current cosmological distance measurements, including 112 high-quality localized fast radio bursts (FRBs), baryon acoustic oscillation (BAO) measurements from the Dark Energy Spectroscopic Instrument Data Release~2 (DESI-DR2) and the Baryon Oscillation Spectroscopic Survey Data Release~12 (BOSS-DR12), Type~Ia supernovae (SNe~Ia) from the PantheonPlus compilation and the Dark Energy Survey Year~5 (DESY5) sample, cosmic chronometers (CC), and the angular scale of the first acoustic peak of the cosmic microwave background (CMB) from Planck~2018. These datasets allow us to effectively break parameter degeneracy, obtain stringent cosmological constraint results, and conduct systematic model comparison and selection. By using the FRB+PantheonPlus+DESI+CC+CMB dataset, we constrain the parameters of the dark energy equation of state in the $w_{0}w_{a}$CDM model, obtaining $w_{0} = -0.867 \pm 0.060$ and $w_{a} = -0.35^{+0.29}_{-0.25}$.
For the $f(R)$ modified gravity models, the deviation parameter $b$, which characterizes departure from general relativity, is constrained to be $b = 0.206 \pm 0.084$, $b = 0.700^{+0.190}_{-0.130}$, and $b = 0.198 \pm 0.081$ for Hu-Sawicki, Starobinsky, and ArcTanh models, respectively. Besides, we compare the impacts of different SNe Ia datasets (PantheonPlus and DESY5) and BAO datasets (DESI-DR2 and BOSS-DR12) on the constraints of the cosmological models.
By employing Bayesian evidence and other model selection criteria, we find that the choice of SNe Ia and BAO datasets can significantly influence the inferred preference for cosmological models. Specifically, the DESY5 and DESI datasets tend to favor $w_{0}w_{a}$CDM and $f(R)$ models, whereas the PantheonPlus and BOSS datasets show a comparatively stronger preference for the $\Lambda$CDM model.
\end{abstract}

\keywords{{Dark energy} --- {Modified gravity} --- {Cosmology} --- {Cosmological model}}

\section{Introduction}
Over the past two decades, one of the most significant achievements in cosmology has been the discovery that the Universe is undergoing accelerated expansion. Since the pioneering observations of Type Ia supernovae (SNe Ia) first revealed this phenomenon \citep{SupernovaSearchTeam:1998fmf,SupernovaCosmologyProject:1997zqe,SupernovaSearchTeam:1998bnz}, its existence has been firmly established by a variety of independent probes, most notably the anisotropies in the cosmic microwave background (CMB) \citep{2007ApJS..170..377S} and measurements of baryon acoustic oscillation (BAO) \citep{2005ApJ...633..560E}. The accelerated expansion of the Universe can be naturally explained by introducing the cosmological constant, i.e. $\Lambda$, into Einstein field equations of general relativity. When combined with cold dark matter (CDM), this framework constitutes the standard cosmological model, i.e. $\Lambda$CDM. 

Although the $\Lambda$CDM model has achieved remarkable success, it is also accompanied by a number of tensions, such as the Hubble tension and the $S_8$ tension \citep{2021CQGra..38o3001D,2022NewAR..9501659P,2023ARNPS..73..153K} , which may hint at new physics beyond the standard model. In this context, dark energy may not be a simple cosmological constant but rather a dynamical component, whose evolution is most commonly parameterized by the Chevallier-Polarski-Linder (CPL) form, expressed as $w(a) = w_0 + w_a (1-a)$ \citep{2001IJMPD..10..213C, 2003PhRvL..90i1301L}. Recent observations further support the need for such a flexible description, as the latest BAO measurements by the DESI collaboration, combined with SNe Ia and CMB data, show a pronounced preference for the $w_0w_a$CDM model \citep{2025arXiv250314738D}.

Alternatively, the current observational data can also be interpreted within the framework of modified gravity.
In 1980, \citet{1980PhLB...91...99S} proposed a modified gravity model, $f(R) = R + \alpha R^2$, to account for the inflationary phenomenon. Since then, $f(R)$ theory, which generalize the Ricci scalar $R$ to an arbitrary function $f(R)$, has become one of the most extensively studied modified gravity models.
Although many $f(R)$ modified gravity models have been proposed, only a subset are considered viable. Some well-known examples include the Hu-Sawicki model \citep{Hu:2007nk}, Starobinsky model \citep{Starobinsky:2007hu}, Tsujikawa model \citep{2008PhRvD..77b3507T}, and the Exponential model \citep{2009PhRvD..80l3528L}.

To facilitate more efficient constraints on $f(R)$ theory from cosmological observations, \cite{Basilakos:2013nfa} developed a method to solve the ordinary differential equations (ODEs) using series expansions, allowing the Hubble parameter to be computed in a form that deviates from $\Lambda$CDM by a deviation parameter $b$. Moreover, this Taylor series approach was subsequently extended to incorporate the ArcTanh model in \cite{2022MNRAS.514.5827S}, which was first proposed by \cite{P_rez_Romero_2018}. With the emergence of cosmological tensions, $f(R)$ theory has become a promising approach, and its constraints have grown increasingly important. Various cosmological observations have been employed to place stringent limits on $f(R)$ modified gravity models \citep{2024MNRAS.527.7626R,2023PDU....4201281K,2024RAA....24k5013Y}.

Since dark energy or modified gravity can change the expansion history of the Universe, it is important to accurately measure the cosmological distance at different redshifts to validate these models. In particular, new methods of measuring cosmological distance are essential for breaking parameter degeneracies, precisely constraining cosmological parameters, and distinguishing theoretical models. 

As an example, Fast radio bursts (FRBs) are extremely bright radio transients first discovered by \cite{2007Sci...318..777L}. Although the origin of FRBs is not yet fully understood \citep{2023RvMP...95c5005Z}, they are already widely discussed and used in cosmology. The frequency-dependent delay of FRB signals, caused by dispersion in the ionized medium, is quantified by the dispersion measure (DM), which traces the integrated free electron density along the line of sight.  
The relation between DM and redshift makes FRB a promising cosmological probe, offering a powerful means to constrain cosmological parameters \citep[e.g.][]{2014ApJ...783L..35D,2014PhRvD..89j7303Z,2024arXiv241024072G}. With an increasing number of FRB host galaxies being localized, FRB cosmology has matured as a tool to probe important cosmological parameters, including the baryon density \citep{2020Natur.581..391M,2023ApJ...944...50W,2022ApJ...940L..29Y,2025arXiv250706841Z}, the Hubble constant \citep{2025A&A...698A.215G,2025ApJ...988..177X,2025PDU....4801926K}, and the dark energy equation of state \citep{2025ApJ...981....9W}.

In this work, we use a sample of 112 high-quality localized FRBs and other cosmological distance-related measurements, i.e. SNe Ia, BAO, cosmic chronometers (CC), and CMB, to constrain and compare the $\Lambda$CDM, $w_0w_a$CDM and $f(R)$ modified gravity models. The paper is organized as follows. In Section~\ref{sec:models}, we introduce the $w_0w_a$CDM model and three representative $f(R)$ modified gravity models. In Section~\ref{sec:Method}, we describe the datasets used and the construction of the likelihood functions. In Section~\ref{sec:results}, we present the constraint results of the model parameters and the results of model comparisons. In Section~\ref{sec:conclusion}, we summarize our results.

\section{models}\label{sec:models}
\subsection{$w_0w_aCDM$ model}
If general relativity provides an accurate description of the dynamical evolution of the Universe, then in a spatially flat universe ($\Omega_\text{K} = 0$), the Hubble parameter $H(z)$ can be written as
\begin{equation}
\begin{aligned}
H(z) = H_0\left[ \Omega_{\text{m}} (1 + z)^3 \right. &+ \Omega_\text{r} (1 + z)^4 \\
&\left. + (1-\Omega_{\text{m}}-\Omega_{\text{r}})f_{\rm DE}(z) \right]^{1/2}\ ,\label{eq1}
\end{aligned}
\end{equation}
where $\Omega_\mathrm{m}$ and $\Omega_\mathrm{r}$ denote the present energy density fractions of matter and radiation, respectively. The radiation energy density fraction is determined by the CMB temperature $T_{\rm CMB}$ and
the effective number of relativistic species $N_{\rm eff}$.
The total radiation density parameter can be written as
\begin{equation}
\Omega_r h^{2} = \Omega_\gamma h^{2}
\left[1 + \frac{7}{8}\left(\frac{4}{11}\right)^{4/3} N_{\rm eff}\right],
\end{equation}
where $\Omega_{\gamma}$ represents the energy density parameter of
photons at present.
We adopt $N_{\rm eff}=3.04$ and $T_{\rm CMB}=2.7255\,\mathrm{K}$ from \cite{2009ApJ...707..916F}, which gives
$
\Omega_\gamma h^{2} \simeq 2.47\times10^{-5},
$ and 
$\Omega_r h^{2} \simeq 4.18\times10^{-5}$.
The $H_0$ is the Hubble constant, and $f_\mathrm{DE}$ characterizes the redshift evolution of the dark energy density governed by its equation of state $w(z)$, which can be expressed as
\begin{equation}
f_{\rm DE}(z) = \exp\left[ 3 \int_0^z \frac{1 + w(z')}{1 + z'} \, dz' \right] \ .
\end{equation}

In the $\Lambda$CDM model, dark energy is assumed to be a cosmological constant, such that $f_{\rm DE} = 1$, independent of redshift. Under the CPL parameterization \citep{2001IJMPD..10..213C, 2003PhRvL..90i1301L}, where $w(z) = w_0 + w_a (1-a) = w_0 + w_a\, z/(1+z)$, the dark energy evolution becomes
\begin{equation}
f_{\rm{DE}}(z) = (1+z)^{3(1 + w_0 + w_a)} e^{-3 w_a\, \frac{z}{1+z}}\ .
\end{equation}
This parameterization can capture the evolution of the dark energy density with redshift for different dark energy models.
When $w_0 = -1$ and $w_a = 0$, it reduces to the $\Lambda$CDM, and thus the $w_0w_a$CDM framework provides a useful means to quantify potential deviations from the standard cosmological model.

\subsection{f(R) theory}
The $f(R)$ modified gravity theory provides a natural generalization of general relativity, in which the Ricci scalar $R$ in the Einstein-Hilbert action is replaced by an arbitrary function $f(R)$ \citep{2011PhR...505...59N}, and we have
\begin{equation}
S = \frac{1}{16 \pi G}  \int d^4 x \, \sqrt{-g} \ f(R)  +S_m + S_r \ ,
\end{equation}
where $S_m$ and $S_r$ denote the actions of matter and radiation, respectively.
By varying the action with respect to the metric, we obtain the field equation as
\begin{equation}
\begin{aligned}
   f_R G_{\mu\nu} - \frac{1}{2} g_{\mu\nu} f &+ \frac{1}{2} g_{\mu\nu} f_R R \\&- \nabla_\mu \nabla_\nu f_R + g_{\mu\nu} \square f_R = 8\pi G T_{\mu\nu} \ .
\end{aligned}
\end{equation}
Here, $G_{\mu\nu} \equiv R_{\mu\nu} - \tfrac{1}{2} g_{\mu\nu} R$ denotes the Einstein tensor, $f \equiv f(R)$ is a general function of the Ricci scalar, and $f_R \equiv df/dR$ represents its derivative with respect to $R$. The operator $\nabla_\mu$ denotes the covariant derivative, $\square \equiv g^{\mu\nu}\nabla_\mu \nabla_\nu$ represents the d'Alembertian operator, and $T_{\mu\nu}$ is the energy-momentum tensor. 

In a spatially flat universe, adopting the Friedmann-Lemaitre-Robertson-Walker (FLRW) metric, the modified Friedmann equation in $f(R)$ theory can be written as
\begin{align}
3 f_R H^2 &= 8\pi G (\rho_m + \rho_r) + \frac{1}{2} (f_R R - f) - 3 H \dot{f}_R \label{eqH_2}\ , \\
-2 f_R \dot{H} &= 8\pi G (\rho_m + p_m + \rho_r + p_r) + \ddot{f}_R - H \ddot{f}_R\ .
\end{align}
Here, the overdot denotes differentiation with respect to cosmic time $t$, and $(\rho_m, \rho_r)$ and $(p_m, p_r)$ represent the energy density and pressure of matter and radiation, respectively.

However, an arbitrary choice of $f(R)$ may lead to severe conflicts with current cosmological and astrophysical observations, including the absence of a proper matter-dominated era \citep{Chiba:2006jp}, i.e. the so-called matter instability \citep{Faraoni:2006sy}, the failure to satisfy local gravity constraints \citep{Nojiri:2006ww}, and instabilities in cosmological perturbations \citep{Bean:2006up}. Therefore, viable $f(R)$ models are required to satisfy the following conditions \citep{2007PhRvD..75h3504A, 2010LRR....13....3D}:
\begin{equation}
\begin{gathered}
0 < \left(\frac{R f_{RR}}{f_{R}} \right)_{r=-2} < 1\ , \\[6pt]
f_{R} > 0 \ \text{and} \ f_{RR} > 0\ , \ \text{for} \ R \ge R_{0} (>0)\ , \\[6pt]
\lim_{R\to 0} f(R) = R\ , 
\lim_{R\to \infty} f(R) = R - 2\Lambda \ .
\end{gathered}
\end{equation}
Here, $f_{RR} \equiv d^2 f/dR^2$ denotes the second derivative of $f(R)$ with respect to $R$, $R_0$ corresponds to the present value of the Ricci scalar, $\Lambda$ is a constant, and $r$ is defined as $r \equiv -R f_R / f$.
Taking these conditions into account, a general expression for a viable $f(R)$ model can be written as
\begin{equation}\label{eqf(R)}
f(R) = R - 2 \Lambda y(R,b)\ ,
\end{equation}
where $y(R,b)$ describes the deviation of the $f(R)$ model from General Relativity, and the parameter $b$ quantifies the magnitude of this deviation.
In this work, we adopt three viable models: the Hu-Sawicki model \citep{Hu:2007nk}, Starobinsky model \citep{Starobinsky:2007hu}, and ArcTanh model \citep{P_rez_Romero_2018}.

The Hu--Sawicki model can be expressed as  
\begin{equation}\label{eqHS}
    f_{\mathrm{HS}}(R) = R - \frac{2\Lambda}{1 + \left(\tfrac{b\Lambda}{R}\right)^n} \ ,
\end{equation}
where $\Lambda \equiv m^2 c_1 / c_2$, $b \equiv 2 c_2^{\,1-1/n}/c_1$, and $m^2 \equiv \kappa^2 \bar{\rho}_0 / 3$ with $\kappa^2 \equiv 1/(16\pi G)$. 
Here, $\bar{\rho}_0$ denotes the average matter density today, $c_1$ and $c_2$ are dimensionless parameters, and $n=1$ is assumed {\citep{2023PDU....4201281K,2022MNRAS.514.5827S}}. 
Combining Eq.~\eqref{eqf(R)} and Eq.~\eqref{eqHS}, we can identify $y_{\mathrm{HS}}(R,b) = \left[1 + \left(\tfrac{b\Lambda}{R}\right)^n \right]^{-1}$. 
Noting that when $b \to 0$, $f_{\mathrm{HS}}(R) \to R - 2\Lambda$, which corresponds to the standard $\Lambda\text{CDM}$ model.

For the Starobinsky model, $f(R)$ is written as
\begin{equation}\label{eqST}
    f_{\mathrm{ST}}(R) = R - 2\Lambda \left\{1- \left[ 1 + \left(\frac{R}{b \Lambda}\right)^2 \right]^{-n} \right\}\ .
\end{equation}
Here we still assume $n=1$, and in the limit $b \to 0$, it reduces to the standard $\Lambda$CDM model. In Eq.~\eqref{eqST}, since the parameter $b$ always appears as $b^2$, the posterior distribution of $b$ is symmetric with respect to $b=0$. Consequently, the physically relevant quantity that characterizes deviations from the standard cosmological model is the absolute value $|b|$. Therefore, following \citet{2024MNRAS.527.7626R}, we impose the prior $b>0$ throughout this work, without any loss of generality.

Similarly, another $f(R)$ form is prosposed to be \citep{P_rez_Romero_2018}
\begin{equation}\label{eq_PN}
    f(R) = R - \frac{2\Lambda}{1+ b\, p(R,\Lambda)}\ ,
\end{equation}
where $p(R,\Lambda)$ represents a parametrization function. We adopt $p(R,\Lambda) = \mathrm{ArcTanh}(\Lambda/R)$ as a specific choice for this parametrization \citep{P_rez_Romero_2018,2022MNRAS.514.5827S}, which is named as the ArcTanh model:
\begin{equation}\label{eq_AcT}
    f_{\mathrm{ACT}}(R) = R - \frac{2\Lambda}{1 + b\, \mathrm{ArcTanh}(\Lambda/R)}\ .
\end{equation}

In order to calculate the cosmological distance for constraining the $f(R)$ models, we need to estimate $H(z)$. In practice, it is rather challenging to directly solve the modified Friedmann equation in $f(R)$ theory. To overcome this difficulty, \citet{Basilakos:2013nfa}
introduced a perturbative approach, in which $H(z)$ in $f(R)$ theory is expanded around the corresponding solution of the $\Lambda$CDM model. Following this method, Eq.~\eqref{eqH_2} is expanded around $b=0$, and only the first two perturbative terms are retained, striking a balance between accuracy and computational efficiency. 
Accordingly, $H(z)$ can be expressed as
\begin{align}
    H^2_{f(R)}(z) = &H^2_{\Lambda}(z) \notag\\
    &+ b \,\delta H^2_{1,f(R)}(z) + b^2 \,\delta H^2_{2,f(R)}(z) + \mathcal{O}(2),
    \label{eqhz1}
\end{align}
for the Hu--Sawicki and ArcTanh model, and
\begin{align}
    H^2_{f(R)}(z) = &H^2_{\Lambda}(z) \notag\\
    &+ b^2 \,\delta H^2_{1,f(R)}(z) + b^4 \,\delta H^2_{2,f(R)}(z) + \mathcal{O}(2),
    \label{eqhz2}
\end{align}
for the Starobinsky model.
Here, $H_{\Lambda}(z)$ denotes the Hubble parameter in the $\Lambda$CDM model, which corresponds to Eq.~\eqref{eq1} in the case of $f_{\mathrm{DE}} = 1$. Remarkably, this truncated expansion reproduces the numerical solution with a discrepancy of less than $0.01\%$ across a broad redshift range, for realistic values of the deviation parameter $b \sim \mathcal{O}(1)$ \citep{Basilakos:2013nfa}.
The explicit $H(z)$ forms of the three $f(R)$ models can be found in the previous works \citep[e.g.][]{2022MNRAS.514.5827S,2024RAA....24k5013Y}.

\section{Observational data} \label{sec:Method}
\subsection{FRB}
Cosmological information is encoded in the FRB dispersion measure (DM). The observed dispersion measure $\mathrm{DM}_{\mathrm{obs}}$ can be decomposed into four components \citep{2014ApJ...783L..35D}, which is given by
\begin{equation}
\mathrm{DM}_{\mathrm{obs}} = \mathrm{DM}_{\mathrm{MW}} + \mathrm{DM}_{\mathrm{IGM}} + \mathrm{DM}_{\mathrm{halo}} + \frac{\mathrm{DM}_{\mathrm{host}}}{1 + z}, \label{eq:DM}
\end{equation}
where $\mathrm{DM}_{\mathrm{MW}}$, $\mathrm{DM}_{\mathrm{IGM}}$, $\mathrm{DM}_{\mathrm{halo}}$, and $\mathrm{DM}_{\mathrm{host}}$ represent the contributions from the Milky Way interstellar medium, the intergalactic medium (IGM), galactic halo, and the host galaxy, respectively. 
The $\mathrm{DM}_{\mathrm{MW}}$ can be estimated using electron density models such as NE2001\citep{2002astro.ph..7156C} and YMW16\citep{2017ApJ...835...29Y}. 
The halo term $\mathrm{DM}_{\mathrm{halo}}$ is typically assumed to lie in the range of $50$--$80~\mathrm{pc~cm^{-3}}$ \citep{2019MNRAS.485..648P}.  Based on these results, previous studies have typically adopted a fixed value for the halo dispersion measure, for example$\mathrm{DM}_{\mathrm{halo}} = 50~\mathrm{pc~cm^{-3}}$\citep[e.g.][]{2020Natur.581..391M,2022MNRAS.516.4862J,2023MNRAS.526.1773F},while other works have used a slightly larger value of
$\mathrm{DM}_{\mathrm{halo}} = 65~\mathrm{pc~cm^{-3}}$
\citep[e.g.][]{2023ApJ...946L..49L,2025ApJ...981....9W}. In this work, we adopt the NE2001 model to compute $\mathrm{DM}_{\mathrm{MW}}$ and take $\mathrm{DM}_{\mathrm{halo}} = 65~\mathrm{pc~cm^{-3}}$, and our results are not sensitive to this value.

For $\mathrm{DM}_{\mathrm{host}}$, since it depends on the properties of the host galaxy and the location of the FRB within it, it remains highly challenging to determine accurately. \citet{2020Natur.581..391M} suggested that a log-normal distribution provides a reasonable   description of $\mathrm{DM}_{\mathrm{host}}$, which can be written as
\begin{equation}
p_{\rm host}(\rm DM_{host}) = \frac{1}{\sqrt{2\pi}{\rm DM}\sigma_{\rm host}}
\\
{\rm exp} \left[ - \frac{({\rm ln \, DM} - \mu)^2}{2\sigma_{\rm host}^2} \right],
\end{equation}
Here we have taken into account the cosmological redshift correction to ${\rm DM_{host}}$.
The distribution has a median of $\mathrm{e}^{\mu}$ and a standard deviation given by $\mathrm{e}^{\mu+\sigma^2_{\rm host}/2}(\mathrm{e}^{\sigma^2_{\rm host}}-1)^{1/2}$. To estimate the probability distribution of $\mathrm{DM}_{\text{host}}$, \cite{2020ApJ...900..170Z} classified host galaxies of FRBs into three categories based on the type of FRB (non-repeating or repeating), host galaxy stellar mass, and star formation rates (SFRs): 
(1) FRB20121102A-like repeating bursts hosted by dwarf galaxies, with stellar masses $M\sim 1-50 \times 10^7 \, M_{\odot}$ and SFRs $\sim 0.1-0.7 \, M_{\odot} \, \rm yr^{-1}$; 
(2) FRB20180916B-like repeating bursts hosted by spiral galaxies, with stellar masses $M\sim 0.1-10 \times 10^{10} \, M_{\odot}$ and SFRs $\sim 0.01-10 \, M_{\odot} \, \rm yr^{-1}$; (3) non-repeating  bursts. Utilizing the IllustrisTNG simulation {\citep{2018MNRAS.480.5113M,2018MNRAS.475..648P,2018MNRAS.477.1206N,2018MNRAS.475..676S,2018MNRAS.475..624N}}, \cite{2020ApJ...900..170Z} provided the preferred values of the parameters $\mu$ and $\sigma_{\rm host}$ at several redshifts. We adopt a cubic spline interpolation to obtain the values of each parameter at a given redshift.

In Eq.~(\ref{eq:DM}), $\mathrm{DM}_{\text{IGM}}$ is closely related to the cosmological parameters, and the mean value of $\mathrm{DM}_{\text{IGM}}$ can be estimated by \citep{2014PhRvD..89j7303Z}
\begin{equation}\label{eq:3}
\langle {{\rm DM_{IGM}}(z)} \rangle = \frac{3 c  \Omega_b H_0 }{8\pi G m_p} \int^{z}_0 \frac{f_{\rm IGM}(z') \chi_e(z')(1+z') {\rm d}z'}{E(z')},
\end{equation}
where
\begin{equation}
\chi_e(z') = Y_{\mathrm{H}} X_{e,\mathrm{H}}(z') + \tfrac{1}{2} Y_{\mathrm{He}} X_{e,\mathrm{He}}(z').
\end{equation}
Here, $m_p$ denotes the proton mass,  $\chi_{e}(z')$ is the fraction of free electrons in the IGM, and $E(z')=H(z')/H_0$. We adopt $Y_{\mathrm{H}} = 3/4$ and $Y_{\mathrm{He}} = 1/4$ as the mass fractions of hydrogen and helium, respectively. For $z<3$, both hydrogen and helium in the IGM are assumed to be fully ionized, such that $X_{e,\mathrm{H}}(z') \simeq  X_{e,\mathrm{He}}(z') \simeq  1$, giving $\chi_{e}(z') \simeq 7/8$. The fraction of baryons residing in the IGM, denoted as $f_{\mathrm{IGM}}$, is still poorly constrained as a function of redshift and is usually treated as a constant. \citet{2012ApJ...759...23S} estimated $f_{\mathrm{IGM}} = 0.84$, whereas \citet{2025NatAs.tmp..131C}, using data from the Deep Synoptic Array (DSA-110), provided a more precise constraint of $f_{\mathrm{IGM}} = 0.93$. In this work, we adopt $f_{\mathrm{IGM}} = 0.93$ \citep{2025A&A...698A.215G,2025ApJ...988..177X}. As we tested, the effect of this value on the constraint results of the dynamical dark energy and $f(R)$ modified gravity models are negligible.

Since the electron number density along the line of sight is not uniform in the IGM, we adopt a quasi-Gaussian distribution with a long tail to describe it \citep{2020Natur.581..391M}, which is obtained by combining observational measurements with simulation results. It is expressed as 
\begin{equation}
p_{\rm IGM}(\Delta)=A\Delta^{-\beta}{\rm exp} \left[ -\frac{(\Delta^{-\alpha} - C_0)^2}{2\alpha^2\sigma_{\rm DM}^2}\right], \quad \Delta>0.
\end{equation}
Here $\Delta = {\rm DM}_{\rm IGM}/\langle {\rm DM}_{\rm IGM} \rangle$ and $\alpha = \beta = 3$ characterize the inner density profile of gas in halos, and $\sigma_{\rm DM}$ denotes the effective standard deviation. $A$ denotes the amplitude, and $C_0$ is a free parameter determined by enforcing $\langle \Delta \rangle = 1$. 
%\cite{2021ApJ...906...49Z} obtained the best-fit values of $A$, $C_0$, and $\sigma_{\rm DM}$ at different redshifts using the state-of-the-art IllustrisTNG simulation.  
\citet{2020Natur.581..391M} modeled the scatter of the IGM dispersion measure as
$\sigma_{\rm DM} = F z^{-0.5}$,
where the parameter $F$ characterizes the strength of baryonic feedback.
However, \citet{2025arXiv250805161Z} pointed out that this functional form is only valid at relatively high redshifts, and that the inclusion of low-redshift data tends to favor larger values of $F$, thereby introducing a systematic bias in the inferred results. To avoid this issue,
we adopt the results from \cite{2021ApJ...906...49Z}, derived by using the state-of-the-art IllustrisTNG simulation {\citep{2018MNRAS.480.5113M,2018MNRAS.475..648P,2018MNRAS.477.1206N,2018MNRAS.475..676S,2018MNRAS.475..624N}}, and employ a cubic spline interpolation to obtain $A$, $C_0$, and $\sigma_{\rm DM}$ at arbitrary redshifts.

\begin{figure}[t!]
	\centering
	\includegraphics[width=\linewidth]{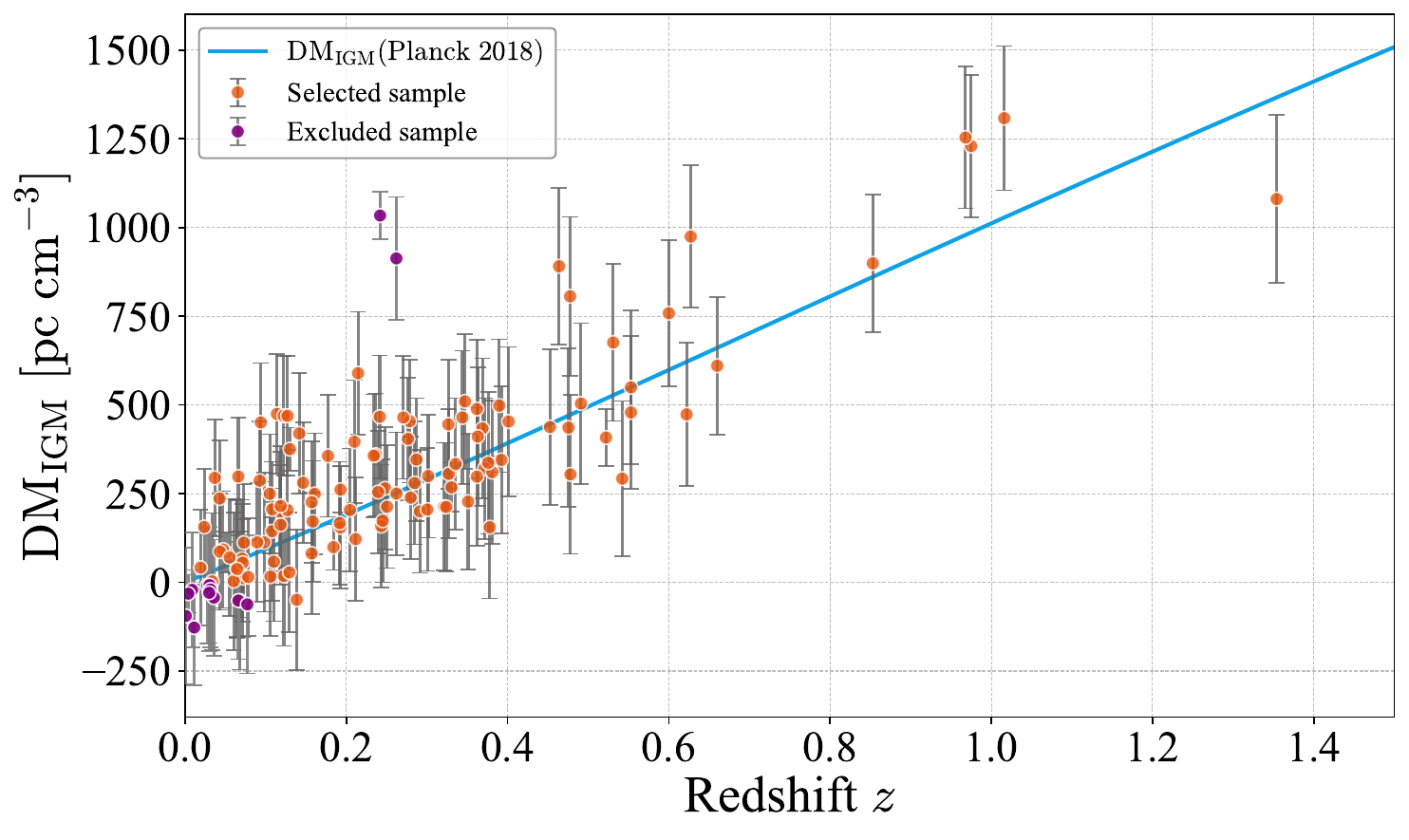}
	\caption{The $\mathrm{DM}_{\mathrm{IGM}}$ as a function of redshift for the localized FRBs. The orange points correspond to the 112 selected FRBs from the 125 localized FRBs, while the 13 purple points represent those excluded due to not satisfying the selection criteria. The blue curve denotes the theoretical calculation from the results of Planck 2018 \citep{2020A&A...641A...6P} assuming the flat $\Lambda$CDM cosmology.} 
	\label{fig:DM_vs_z}
\end{figure}

Considering the probability distribution from both $p_{\rm host}$ and $p_{\rm IGM}$, the likelihood function of all FRBs can be calculated as
\begin{align}
\mathcal{L}_{\rm FRBs} &= \prod^{N}_{i} \int^{{\rm DM}_{{\rm ex},\,i}}_{0} \frac{1}{\langle {\rm DM_{IGM}} \rangle} \, p_{{\rm host},\,i}({\rm DM}_{\rm host}) \notag \\
&\quad \times p_{{\rm IGM},\,i}(\Delta) \, {\rm dDM}_{\rm host}\ ,
\end{align}
where ${\rm DM}_{{\rm ex},\,i}={\rm DM}_{{\rm obs},\,i} - {\rm DM}_{{\rm MW},\,i} - {\rm DM}_{{\rm halo},\,i}$
denotes the extragalactic dispersion measure of the $i$th FRB. \cite{2025arXiv250406845Z} and \cite{2025arXiv250805161Z}  found that a correction factor $1/\langle {\rm DM}_{\rm IGM} \rangle$ should be included, since the variable in $p_{\rm IGM}$ is actually $\Delta$ rather than ${\rm DM}_{\rm IGM}$. To avoid the impact of IGM fluctuations at low redshift, the peculiar velocities of host galaxies and unreasonable values for ${\rm DM}_{\rm ex}$ \citep{2025ApJ...981....9W,2025A&A...698A.215G,2025arXiv250805161Z}, we adopt the selection criterion 
$\rm DM_{obs} - DM_{MW} > 100~{\rm pc~cm^{-3}}$, which can yield more robust results. In addition, we exclude FRB20190520B and FRB20220831A, as these two events exhibit significantly enhanced dispersion measures due to their extreme local environments. Finally, we select 112 FRBs out of 125 localized events as our sample, as shown in Figure~\ref{fig:DM_vs_z}. The measurement uncertainties arise from three components: $\mathrm{DM}_{\mathrm{MW}}$ estimated with the NE2001 model, with an uncertainty of about $30~\mathrm{pc~cm^{-3}}$; a halo contribution in the range $50$--$80~\mathrm{pc~cm^{-3}}$, with an uncertainty of $15~\mathrm{pc~cm^{-3}}$; and a host galaxy contribution, whose value and uncertainty are determined by interpolation according to the host galaxy type. The details of this FRB sample are listed in Table~\ref{Tab:FRBs} in Appendix~\ref{app:data}.

\subsection{SNe Ia}
SNe Ia are triggered when the mass of a white dwarf exceeds the Chandrasekhar limit, at which stage the electron degeneracy pressure can no longer resist gravitational collapse.
Since the explosion occurs at nearly the same mass, they can serve as standard candles, providing robust measurements of cosmological distances. We employ two distinct SNe Ia datasets, PantheonPlus\footnote{\url{https://github.com/PantheonPlusSH0ES/DataRelease}}\citep{2022ApJ...938..110B,2022ApJ...938..113S,2022ApJ...938..111B,2022ApJ...938..112P,2022PASA...39...46C,2023ApJ...945...84P,2023ApJ...944..188B} and DESY5\footnote{\url{https://github.com/des-science/DES-SN5YR}}\citep{2024ApJ...975....5S,2024ApJ...973L..14D,2024ApJ...975...86V}, for cosmological analysis. The PantheonPlus compilation consists of 1701 light curves from 1550 distinct SNe Ia, spanning the redshift range $0.001 < z < 2.26$ \citep{2022ApJ...938..110B}. All of these SNe Ia have spectroscopic confirmation, ensuring reliable classification of both supernova type and redshift. To mitigate the impact of peculiar velocities at very low redshift, we restrict our analysis to $z > 0.01$, resulting in a working sample of 1590 light curves. In contrast, the DESY5 sample includes 1635 DES SNe in the redshift range $0.10 < z < 1.13$, together with an external low-$z$ sample of 194 SNe Ia with $0.01 < z < 0.10$ \citep{2024ApJ...973L..14D}. In total, the final DESY5 dataset comprises 1829 SNe Ia.  Unlike PantheonPlus, which is spectroscopically classified, the DESY5 sample is photometrically classified. 

The two datasets also differ in their treatment in cosmological analysis. In SN~Ia cosmology, the absolute magnitude $M$ and the Hubble constant $H_0$ cannot be independently constrained, as they are fully degenerate. Accordingly, the PantheonPlus sample provides the observed $B$-band peak magnitude $m_B$ together with the uncertainty, so that the absolute magnitude $M$ can be treated as a free parameter to be constrained \citep{2022ApJ...938..110B}. 
By comparison, the DESY5 introduces the nuisance parameter $\mathcal{M} \equiv M + 5\log_{10}(c/H_0)$, which combines the dependence on $M$ and $H_0$, and then marginalizes over $\mathcal{M}$ \citep{2024ApJ...973L..14D}. Therefore, when using the DESY5 data, $M$ does not need to be treated as a free parameter and no constraint on $M$ is obtained. Finally, we will use these two datasets separately in the fitting process, and discuss the differences of their corresponding constraint results.

In the $\Lambda \rm CDM$ or $w_0w_a\rm CDM$ cosmological model, the distance modulus $\mu_{\rm DE}$ of SNe Ia can be derived as
\begin{equation}
    \mu_{\rm DE}=m_{B}-M = 5 \log_{10}\left(\frac{d_L}{\rm Mpc}\right) + 25\ .
\end{equation}
Here, $d_L(z)$ is the luminosity distance at redshift $z$, and in a flat universe, it can be written as
\begin{equation}
    d_L(z) = (1+z)\int^z_0 \frac{c}{H(z')} d z'\ .
\end{equation}

On the other hand, in $f(R)$ gravity, the Newtonian gravitational constant $G$ is no longer strictly constant.  Consequently, a temporal variation of $G$ induces a redshift dependence in the average rescaled intrinsic peak luminosity of SNe~Ia, denoted as $G_{f(R)}(z)$, which can be expressed as \citep{2007PhRvD..76b3514T} 
\begin{equation}\label{eq30}
    G_{f(R)}(z) = \frac{G}{f_R} \left(  \frac{1 + 4k^2 m /a^2}{1 + 3k^2m/a^2} \right)\ ,
\end{equation}
where $m = f_{RR}/f_{R}$ and $k = 0.1\ h\, {\rm Mpc^{-1}}$. 
As a result, the intrinsic luminosity of SNe~Ia, which is proportional to the Chandrasekhar mass $M_{\rm Ch} \propto G^{-3/2}$, can be modified. It has been shown that the light-curve width--luminosity relation continues to hold even when $G \neq G_0$~\citep{2018PhRvD..97h3505W}, 
%In this case, light curves rescaled to match the shape around the peak still exhibit a good agreement in their overall profiles, albeit with a slightly increased scatter. 
but the resulting rescaled intrinsic peak luminosities systematically differ from the local template calibrated at $G = G_0$. Therefore, we only need to consider the effect of a varying gravitational strength on the distance modulus, which is modified in $f(R)$ gravity as \citep{2001PhRvD..65b3506G,2018PhRvD..97h3505W}
\begin{equation}\label{eq31}
    \mu_{f_{(R)}} = 5 \log_{10}\left(\frac{d_L}{\rm Mpc}\right) + 25 + \frac{15}{4} \log_{10}\left(\frac{G_{f(R)}(z)}{G}\right)\ .
\end{equation}
By fixing $b=0.1$ in the calculations of \cite{2023PDU....4201281K}, it is found that $G_{f(R)}(z)/G - 1$ is at the level of $\sim 10^{-4}$ in the Hu--Sawicki model and of order $\sim 10^{-11}$ in the Starobinsky model, and we find that it is also at the level of $\sim 10^{-4}$ in the ArcTanh model. Therefore, the effect of $G_{f(R)}(z)/G$ is very small in this analysis.

Then, the chi-square of SN Ia can be expressed as
\begin{equation}
\chi^2_{\rm SN} = {\rm \Delta}{\boldsymbol{\mu}}^{T} \boldsymbol{C}_{\rm SN}^{-1} {\rm \Delta}{\boldsymbol{\mu}} \ .
\end{equation}
Here ${\rm \Delta}\boldsymbol{\mu}$ denotes the vector of residuals in the SN Ia distance modulus, defined as $\Delta \mu_i = \mu_{\mathrm{obs},i} -\mu_{\mathrm{model}}(z_i)$, where $\mu_{\mathrm{obs},i}$ is the observed distance modulus of the $i$-th supernova, and $\mu_{\mathrm{model}}(z_i)$ is the corresponding theoretical prediction at redshift $z_i$. $\boldsymbol{C}_{\rm SN}$ is the covariance matrix of the SN data.

\subsection{BAO}
BAO can serves as a standard ruler in cosmological analysis. We use the measurements of $D_{\rm M}/r_{\rm d}$, $D_{\rm H}/r_{\rm d}$, and $D_{\rm V}/r_{\rm d}$ from DESI-DR2 \citep{2025arXiv250314738D}, which characterize the BAO feature in the transverse, radial, and volume-averaged dimensions, respectively, and $D_{\rm M}\, r_{\rm d}^{\rm fid}/r_{\rm d}$, $D_{\rm H}\, r_{\rm d}^{\rm fid}/r_{\rm d}$ from BOSS-DR12 \citep{2017MNRAS.470.2617A}. The $r_{\rm d}^{\rm fid}=147.78~\mathrm{Mpc}$ is the fiducial sound horizon at the drag epoch \citep{2017MNRAS.470.2617A}. The BAO measurements used in this work are summarized in the Table~\ref{tab:DESI} and Table~\ref{tab:BOSS} presented in Appendix~\ref{app:data}.

In a flat universe, the distances $D_{\rm M}$, $D_{\rm H}$, and $D_{\rm V}$ can be derived by
\begin{equation}\label{eq6}
\begin{aligned}
D_{\rm M}(z) & = \int^{z}_{0} \frac{c \, {\rm d}z'}{H(z')} \ , \\
D_{\rm H}(z) & = \frac{c}{H(z)}\  ,\\
D_{\rm V}(z) & = \left[zD_{\rm M}(z)^2 D_{\rm H}(z) \right]^{1/3}\ .
\end{aligned}
\end{equation}

Here, $D_{\rm M}(z)$ is the comoving angular diameter distance, $D_{\rm H}(z)$ is the Hubble distance, and $D_{\rm V}(z)$ is the volume-averaged distance. The comoving sound horizon at the baryon drag epoch, $r_{\rm d}$, is defined as the maximum distance that acoustic waves can travel in the primordial plasma before baryons decouple from photons. It can be expressed as
\begin{equation}
    r_{\rm d} = \int_{z_{\rm d}}^{\infty} \frac{c_{\rm s}(z)}{H(z)} \, {\rm d}z\  ,
    \label{rd}
\end{equation}
where $z_{\rm d}$ is the redshift of the baryon drag epoch, and $c_{\rm s}(z)$ denotes the sound speed in the primordial plasma, which can be derived by
\begin{equation}
c_\mathrm{s}(z) = \frac{c}{\sqrt{3\left(1 + \dfrac{3\Omega_b}{4\Omega_\gamma} \dfrac{1}{1+z}\right)}} .
\end{equation}
where $\Omega_{\gamma}$ represents the energy density parameter of photons at present. The $r_{\rm d}$ is determined by the cosmological parameters and can be written as \citep{2023JCAP...04..023B}
\begin{equation}
\begin{split}
r_{\rm d}\ &= 147.05\,{\rm Mpc} \times \\
&\quad \left( \frac{\Omega_{b}h^2}{0.02236} \right)^{-0.13}
\left( \frac{\Omega_{m}h^{2}}{0.1432} \right)^{-0.23}
\left( \frac{N_{\rm eff}}{3.04} \right)^{-0.1}\,.
\end{split}
\end{equation}
The chi-square of BAO is given by
\begin{equation}
\chi^2_{\rm{BAO}}=
\Delta\boldsymbol{X}^{T} 
    \boldsymbol{C}_{\rm BAO}^{-1} 
    \Delta\boldsymbol{X} 
\ .
\end{equation}
Here $\Delta\boldsymbol{X}=\boldsymbol{X-\boldsymbol{X}_{\text{th}}}$, 
where $\boldsymbol{X}$ denotes the vector of observed BAO quantities, 
$\boldsymbol{X}_{\text{th}}$ represents the corresponding theoretical predictions from a given cosmological model, 
and $\boldsymbol{C}_{\rm BAO}$ is the covariance matrix of the observational BAO data.

\begin{figure}[t!]
	\centering
	\includegraphics[width=\linewidth]{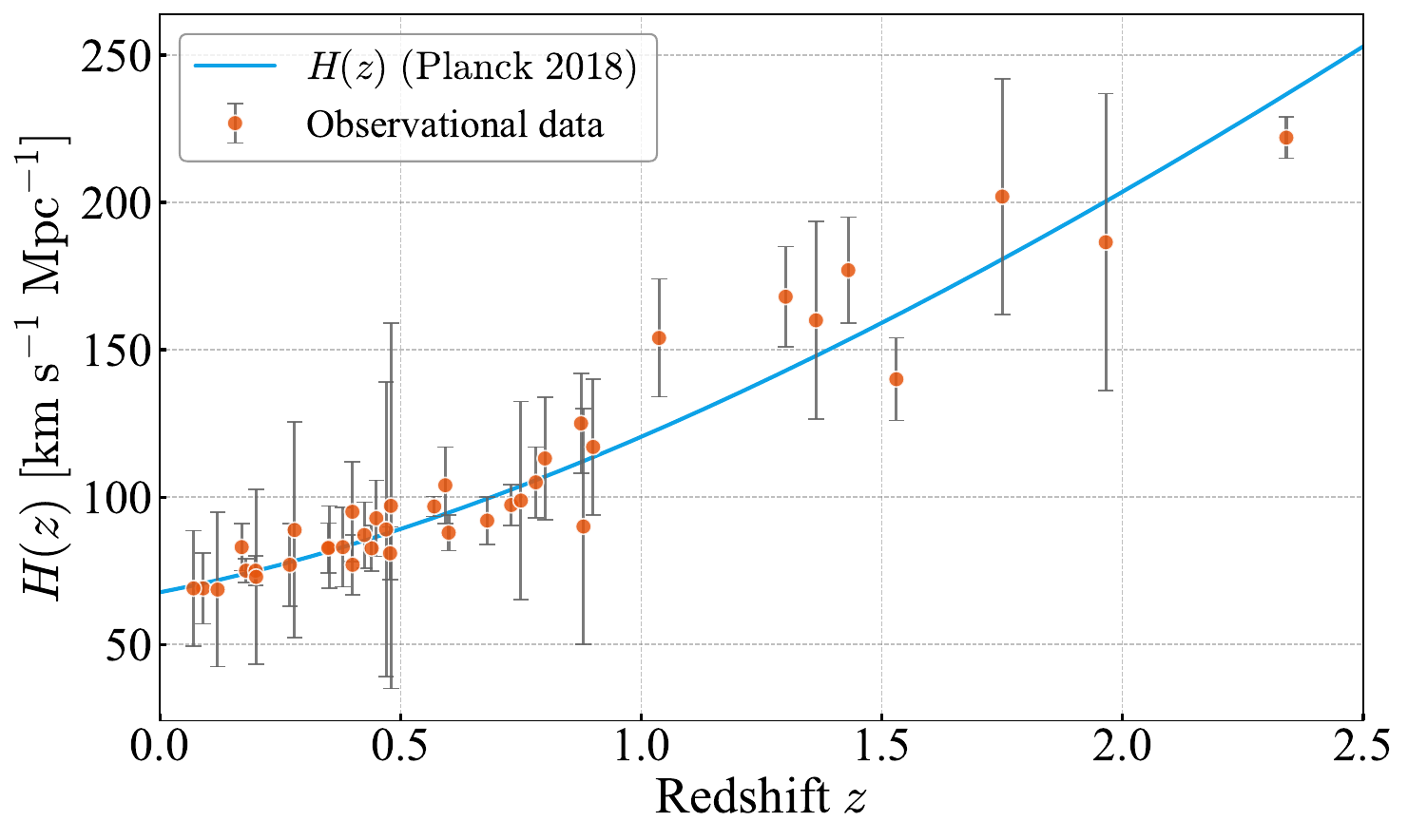}
	\caption{The Hubble parameter $H(z)$ as a function of redshift $z$ based on the cosmic chronometer data we use. The error bars represent the $1\sigma$ CL, which include contributions from both systematic and statistical uncertainties. The blue curve denote the theoretical predictions from Planck 2018 \citep{2020A&A...641A...6P}.}
	\label{fig:CC}
\end{figure}

\subsection{Cosmic Chronometer}
The cosmic chronometer approach can provide a model-independent way to measure the Hubble parameter $H(z)$ \citep{2002ApJ...573...37J}. The key idea of this method is that by estimating the differential age ${\rm d}t$ evolution of the Universe, across a small redshift interval ${\rm d}z$, one can directly infer $H(z)$ without relying on an external cosmological model. The Hubble parameter $H(z)$ can be expressed as
\begin{equation}
H(z) = -\frac{1}{1+z}\,\frac{{\rm d}z}{{\rm d}t}\ .
\end{equation}
In this work, we collect a total of 39 cosmic chronometer data, including the redshift $z$, the Hubble parameter $H(z)$, and 1$\sigma$ uncertainty, which are presented in Figure~\ref{fig:CC} and Table~\ref{tab:cc} in Appendix~\ref{app:data}.

The chi-square of the CC is given by 
\begin{equation}
\chi^2_{\mathrm{CC}} = 
\sum_{i}^{N} 
\frac{\big(H_{\mathrm{obs}}(z_i)-H_{\mathrm{th}}(z_i)\big)^2}
{\sigma_{\text{H},i}^2}\ ,
\end{equation}
where $H_{\mathrm{obs}}(z_i)$ denotes the observational Hubble parameter, $H_{\mathrm{th}}(z_i)$ is the theoretical prediction for a given cosmological model, and $\sigma_{\text{H},i}$ is the corresponding measurement uncertainty.

\subsection{CMB}
The angular scale of the first acoustic peak in the CMB can serve as a crucial observable for measuring cosmic distance, denoted as $\theta_\ast$, which can be written as
\begin{equation}
\theta_\ast = \frac{r_\ast}{D_{\rm M}(z_\ast)}\ ,
\end{equation}
where $z_\ast$ is the redshift of recombination, which depends on the cosmological matter density parameters and can be given by 
\citep{2021PhRvD.104d3521A}
\begin{equation}
\begin{gathered}
z_\ast = \frac{391.672  (\Omega_m h^2)^{-0.372296} + 937.422  (\Omega_b h^2)^{-0.97966}}{(\Omega_m h^2)^{-0.0192951}  (\Omega_b h^2)^{-0.93681}} \\+ (\Omega_m h^2)^{-0.731631}, \
\end{gathered}
\end{equation}
The $r_\ast$ is the comoving sound horizon at $z_\ast$, which can be calculated using Eq.~\eqref{rd} by replacing $z_{\rm d}$ with $z_\ast$.
Compared with the fitting formula in \cite{1996ApJ...471..542H}, which achieves an accuracy of about $0.3\%$, \cite{2021PhRvD.104d3521A} provides a more accurate fitting formula with a precision reaching $0.0005\%$.

We adopt the results from Planck 2018 $(\rm TT,TE,EE+lowE+lensing)$  \citep{2020A&A...641A...6P}, which gives
\begin{equation}
\begin{aligned}
100\theta_\ast &= 1.04110 \pm 0.00031\ ,
\end{aligned}
\end{equation}
The chi-square of the first acoustic peak of the CMB can be expressed as
\begin{equation}
\chi^2_{\text{CMB}}=  \frac{\left( \theta_{\ast,\text{obs}} - \theta_{\ast,\text{th}} \right)^2}{\sigma_{\ast}^2}\ ,
\end{equation}
where $\theta_{\ast,\mathrm{obs}}$ is the observed value, $\theta_{\ast,\mathrm{th}}$ is the theoretical prediction, and $\sigma_{\ast}$ is the uncertainties of $\theta_\ast$.

Finally, considering $\mathcal{L} \propto {\rm exp}(-\tfrac{1}{2}\chi^2)$,
the total likelihood function in a joint analysis can be written as the product of the likelihood function of each dataset:
\begin{equation}
\mathcal{L_\text{total}} = \mathcal{L_\text{FRBs}} \times\mathcal{L_\text{SN}} \times \mathcal{L_\text{BAO}} \times \mathcal{L_\text{CC}} \times \mathcal{L_\text{CMB}}.
\end{equation}
We note that FRBs are tightly sensitive to the parameter combination $\Omega_b H_0$ (see Eq.~(\ref{eq:3})), while cosmic chronometers provide direct measurements of the Hubble parameter $H_0$. As a result, the inclusion of these data helps to break the parameter degeneracies, such as $H_0$, the absolute magnitude of SN~Ia $M$, and the sound horizon scales $r_\mathrm{d}$ and $r_\ast$ that appears in BAO and CMB analyses.

Therefore, our joint dataset has a significant advantage in placing stringent and complementary constraints on each cosmological parameter, by effectively breaking the parameter degeneracies inherent in individual datasets and improving the overall precision of cosmological parameter estimation.
\vspace{-1cm}

\begin{deluxetable}{cc}[t!]
\centering
\tablecaption{The free parameters and priors considered in this work. \label{Tab:priors}}
\setlength{\tabcolsep}{20pt}
\tablehead{\colhead{Parameter} & \colhead{Uniform Prior}}
\startdata
$\Omega_{m}$      & $(0.1,\,0.9)$ \\
$\Omega_{b}$     & $(0.01,\,0.1)$ \\
$H_0$ (km s$^{-1}$ Mpc$^{-1}$) & $(50,\,100)$ \\
$M$ & $(-21,\,-17)$ \\
$w_0$   & $(-3,\,1)$ \\
$w_a$   & $(-5,\,2)$ \\
$b$     & $(-2,\,2) \rm \ \ or \ \ (0,\,2)$ 
\enddata
\end{deluxetable}

\begin{deluxetable*}{ccc@{\hskip 0.8pt}c@{\hskip 0.8pt}c@{\hskip 0.8pt}cc@{\hskip 1pt}c}[ht!]
% \begin{deluxetable*}{cccccccc}[ht!]
\centering
\caption{\label{tab:results} The constraint results of the cosmological parameters for different models using different datasets. HS, ST, and ACT refer to the Hu-Sawicki, Starobinsky, and ArcTanh $f(R)$ modified gravity models, respectively.}
\tablehead{
\colhead{Model} & \colhead{$\Omega_{m}$} & \colhead{$\Omega_{b}$} & \colhead{$H_0$ (km s$^{-1}$ Mpc$^{-1}$)} & \colhead{$b$} & \colhead{$w_0$} & \colhead{$w_a$} & \colhead{$M$}
}
\startdata
\multicolumn{8}{l}{\textbf{Datasets: FRBs+PantheonPlus+DESI +CC+CMB}} \\ [4pt]
$\Lambda$CDM  & $0.297\pm0.005$ & $0.0463\pm0.0010$ & $67.77\pm0.63$ & $-$ & $-$ & $-$ & $-19.43\pm0.02$ \\
$w_0w_a$CDM   & $0.308\pm0.007$ & $0.0482\pm0.0014$ & $66.61\pm0.76$ & $-$ & $-0.867\pm0.060$ & $-0.35^{+0.29}_{-0.25}$ & $-19.45\pm0.02$ \\
HS            & $0.292\pm0.005$ & $0.0464\pm0.0010$ & $68.10\pm0.64$ & $0.206\pm0.084$ & $-$ & $-$ & $-19.46\pm0.02$ \\
ST            & $0.303^{+0.005}_{-0.007}$ & $0.0477\pm0.0012$ & $67.08^{+0.79}_{-0.70}$ & $0.700^{+0.190}_{-0.130}$ & $-$ & $-$ & $-19.45\pm0.02$ \\
ACT           & $0.293\pm0.005$ & $0.0465\pm0.0010$ & $68.02\pm0.64$ & $0.198\pm0.081$ & $-$ & $-$ & $-19.46\pm0.02$ \\
\hline
\multicolumn{8}{l}{\textbf{Datasets: FRBs+DESY5+DESI+CC+CMB}} \\[4pt] 
$\Lambda$CDM  & $0.300\pm0.005$ & $0.0462\pm0.0011$ & $67.49\pm0.64$ & $-$ & $-$ & $-$ & $-$ \\
$w_0w_a$CDM   & $0.317\pm0.007$ & $0.0490\pm0.0014$ & $65.87\pm0.74$ & $-$ & $-0.775\pm0.064$ & $-0.67^{+0.31}_{-0.28}$ & $-$ \\
HS            & $0.292\pm0.005$ & $0.0465\pm0.0010$ & $68.01\pm0.64$ & $0.291\pm0.078$ & $-$ & $-$ & $-$ \\
ST            & $0.311^{+0.006}_{-0.009}$ & $0.0487^{+0.0013}_{-0.0015}$ & $66.21^{+0.97}_{-0.77}$ & $0.900\pm0.140$ & $-$ & $-$ & $-$ \\
ACT           & $0.293\pm0.005$ & $0.0466\pm0.0010$ & $67.90\pm0.64$ & $0.285\pm0.076$ & $-$ & $-$ & $-$ \\
\hline
\multicolumn{8}{l}{\textbf{Datasets: FRBs+PantheonPlus+BOSS+CC+CMB}} \\[4pt]
$\Lambda$CDM  & $0.306\pm0.009$ & $0.0471\pm0.0013$ & $67.17\pm0.80$ & $-$ & $-$ & $-$ & $-19.45\pm0.02$ \\
$w_0w_a$CDM   & $0.306^{+0.012}_{-0.011}$ & $0.0485\pm0.0016$ & $66.30\pm1.00$ & $-$ & $-0.896^{+0.064}_{-0.078}$ & $-0.18^{+0.55}_{-0.39}$ & $-19.46^{+0.04}_{-0.03}$ \\
HS            & $0.292\pm0.011$ & $0.0462\pm0.0013$ & $68.02\pm0.92$ & $0.220\pm0.100$ & $-$ & $-$ & $-19.46\pm0.02$ \\
ST            & $0.304\pm0.009$ & $0.0475\pm0.0013$ & $67.03\pm0.86$ & $0.620^{+0.250}_{-0.130}$ & $-$ & $-$ & $-19.45\pm0.02$ \\
ACT           & $0.293\pm0.010$ & $0.0463\pm0.0013$ & $67.94\pm0.90$ & $0.211\pm0.097$ & $-$ & $-$ & $-19.46\pm0.02$ \\
\hline
\multicolumn{8}{l}{\textbf{Datasets: FRBs+DESY5+BOSS+CC+CMB}} \\[4pt]
$\Lambda$CDM  & $0.313\pm0.009$ & $0.0474\pm0.0013$ & $66.61\pm0.78$ & $-$ & $-$ & $-$ & $-$ \\
$w_0w_a$CDM   & $0.318^{+0.013}_{-0.011}$ & $0.0492\pm0.0015$ & $65.78\pm0.91$ & $-$ & $-0.783^{+0.078}_{-0.100}$ & $-0.68^{+0.68}_{-0.52}$ & $-$ \\
HS            & $0.291\pm0.010$ & $0.0460\pm0.0013$ & $67.98\pm0.93$ & $0.324\pm0.093$ & $-$ & $-$ & $-$ \\
ST            & $0.311^{+0.009}_{-0.011}$ & $0.0485^{+0.0014}_{-0.0016}$ & $66.20^{+1.00}_{-0.84}$ & $0.850^{+0.160}_{-0.130}$ & $-$ & $-$ & $-$ \\
ACT           & $0.292\pm0.010$ & $0.0462\pm0.0013$ & $67.84\pm0.91$ & $0.309\pm0.089$ & $-$ & $-$ & $-$ \\
\enddata
\end{deluxetable*}

\begin{figure*}[ht!]
    \centering
    \includegraphics[width=0.75\textwidth]{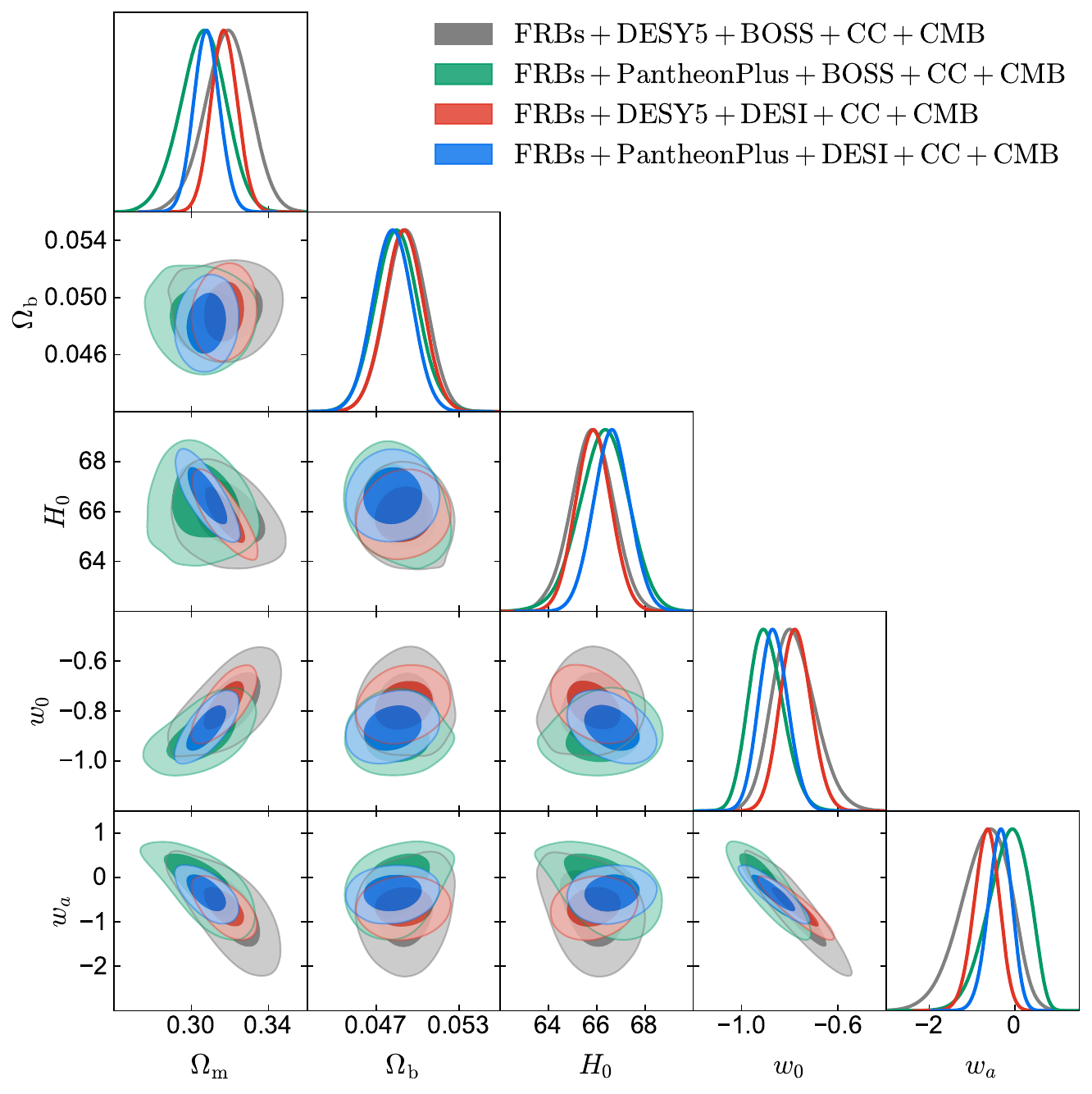}
    \caption{The contour maps (68\% and 95\% CL) and 1D PDFs of $\Omega_m$, $\Omega_b$, $H_0$, $w_0$, and $w_a$ in the flat $w_0w_a$CDM model, derived from different datasets. 
}
    \label{fig:w0wa}
\end{figure*}

\begin{figure}
	\centering
	\includegraphics[width=0.85\linewidth]{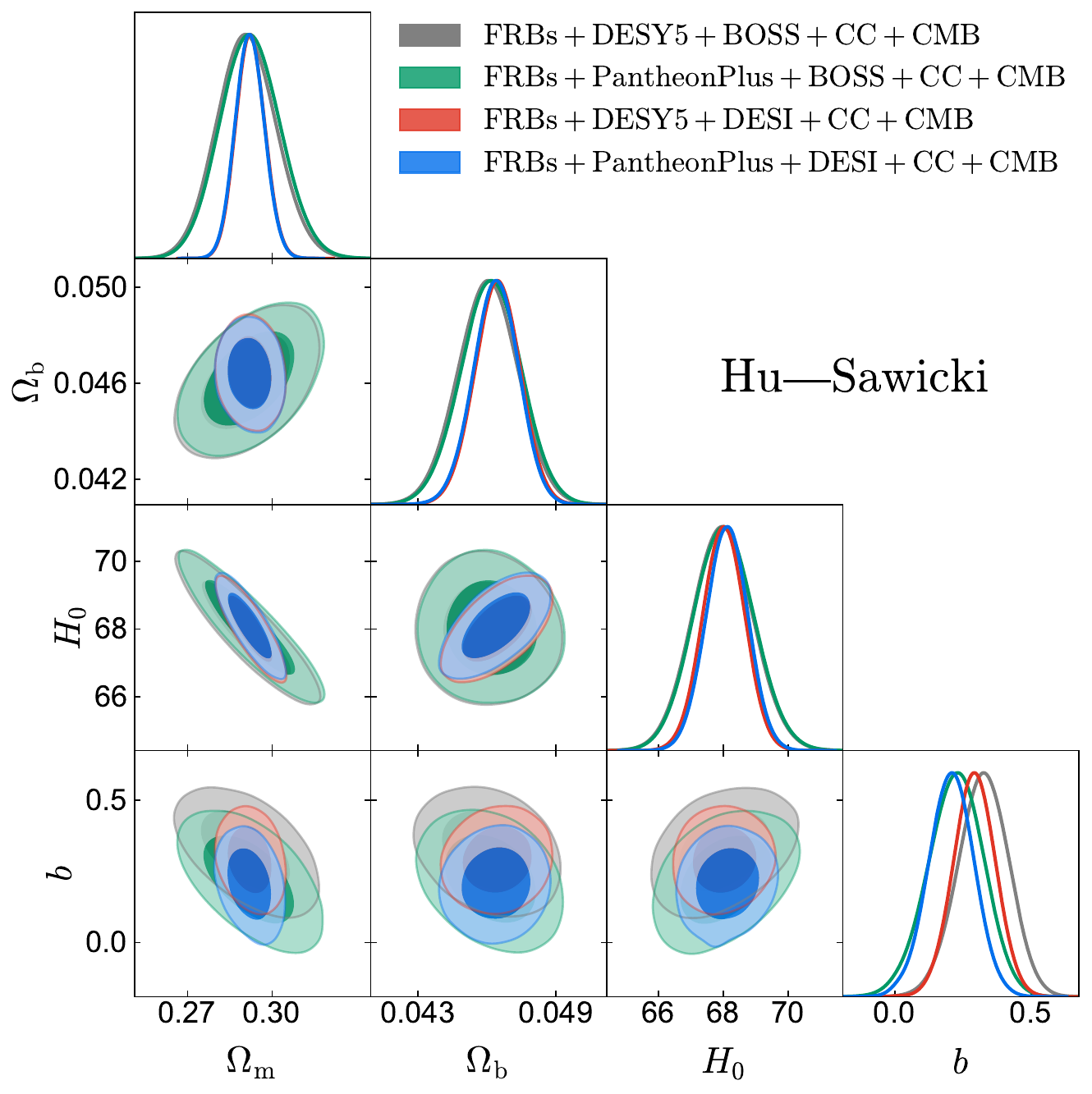}
    \includegraphics[width=0.85\linewidth]{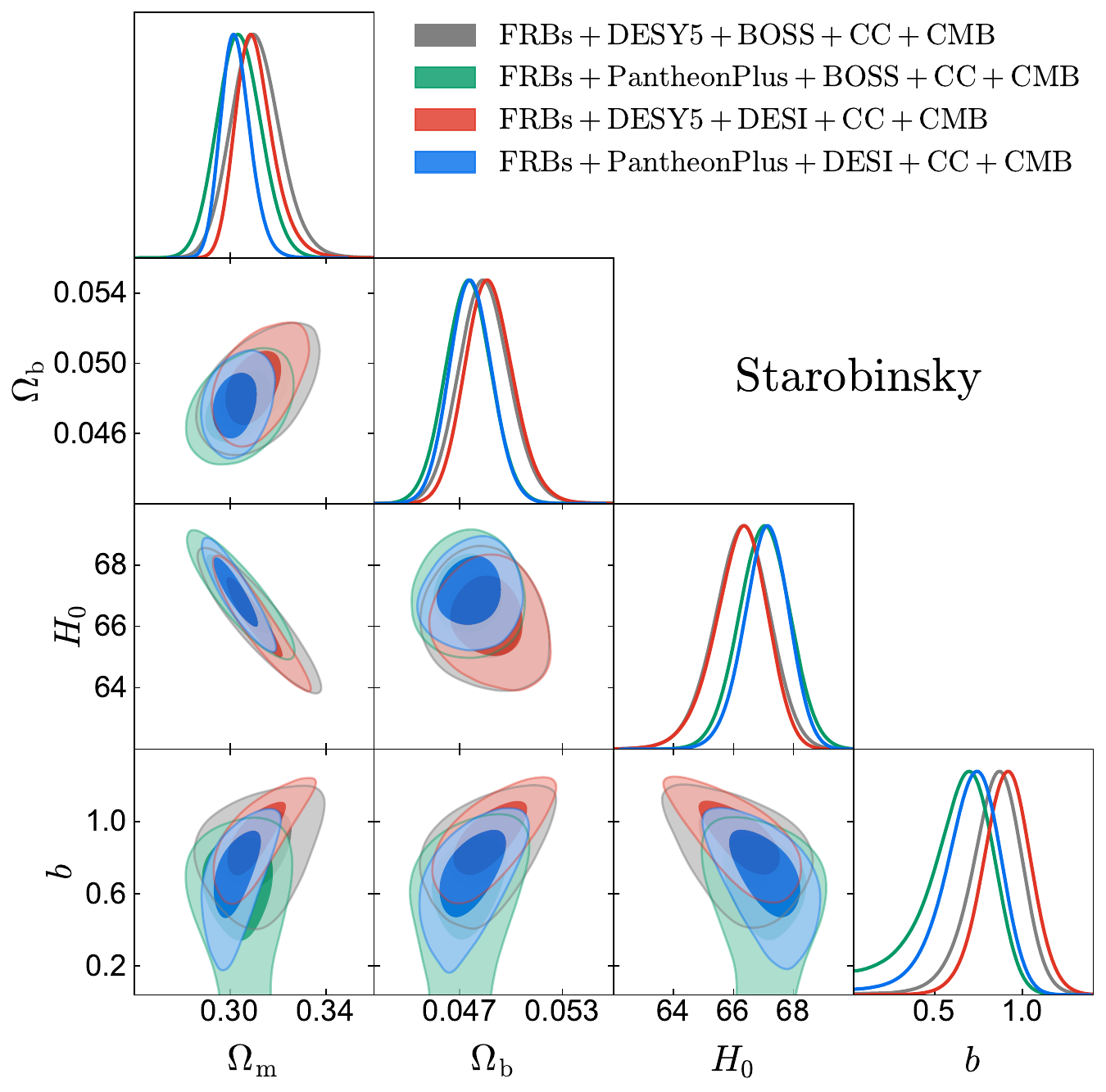}
    \includegraphics[width=0.85\linewidth]{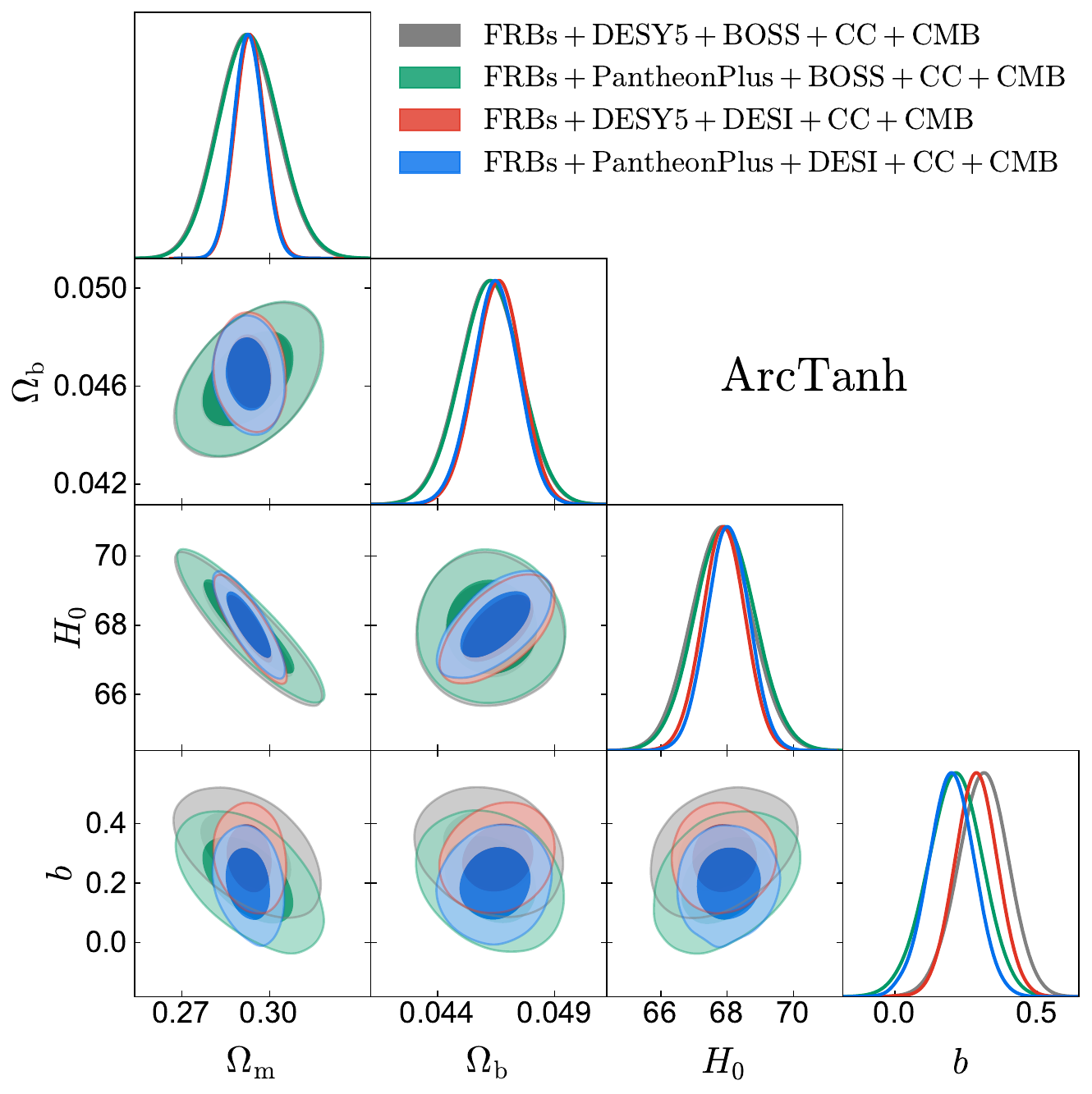}
	\caption{\small  
The contour maps (68\% and 95\% CL) and 1D PDFs of $\Omega_m$, $\Omega_b$, $H_0$, and $b$ for the Hu–Sawicki, Starobinsky, and ArcTanh models (from top to bottom), obtained by fitting different datasets.
}
	\label{fig:fR}
\end{figure}

%\vspace{-1cm}

\section{Constraint and Comparison Results}\label{sec:results}
\subsection{Parameter Constraint}

We constrain the model parameters within the Bayesian framework and perform the Markov Chain Monte Carlo (MCMC) analysis using the \texttt{emcee} \footnote{\url{https://github.com/dfm/emcee}}\citep{emcee} package. The flat priors for all the free parameters are adopted, which are summarized in Table~\ref{Tab:priors}.
The sampling is carried out with 32 independent walkers for $100{,}000$ steps, discarding the first $1{,}000$ steps of each chain as the burn-in process. The remaining samples are then used to construct the posterior probability density functions (PDFs). In Table~\ref{tab:results}, we present the constraint results of the cosmological parameters for different models obtained from the joint datasets, including FRBs, SNe Ia from PantheonPlus or DESY5, BAO from DESI or BOSS, cosmic chronometers, and CMB.
In Figure~\ref{fig:w0wa} and Figure~\ref{fig:fR}, we show the contour maps and 1D PDFs of $\Omega_m$, $\Omega_b$, $H_0$, $w_0$,  $w_a$ and $b$ in the flat $w_0w_a$CDM and $f(R)$ models, obtained from FRBs + PantheonPlus or DESY5 + DESI or BOSS + CC + CMB. 

We find that the constraint results of $w_0$ and $w_a$ of the $w_0w_a$CDM model from all datasets support $w_0 > -1$ and $w_a < 0$, which is consistent with that from \citet{2025arXiv250314738D}. When incorporating the DESI data, our joint dataset yields substantially tighter constraints on $w_0$ and $w_a$, with the precision of $w_a$ in particular improved by about $40\%$ compared to the DESI+PantheonPlus or DESY5 combinations in \citet{2025arXiv250314738D}. Compared with the results of \citet{2025arXiv250314738D} when the full CMB data are further included, the precision of our constraints on both $w_0$ and $w_a$ has also reached a comparable level.

For the three $f(R)$ modified gravity models, i.e. Hu-Sawicki, Starobinsky, and ArcTanh, the joint constraints with different datasets give  consistent results for the constraints on $\Omega_m$, $\Omega_b$, and $H_0$, which are also in agreement with those predicted by the $w_0w_a$CDM model. For the constraint on $b$, all datasets prefer a $b>0$ result for the three $f(R)$ models, especially for the Starobinsky model with a larger value of $|b|\simeq0.6$--$0.9$, whose constraint on $b$ shows a lower precision compared to those of the Hu--Sawicki and ArcTanh models.
Besides, the datasets including DESY5 basically favor larger $b$ values compared to those including PantheonPlus, implying a stronger deviation from the $\Lambda$CDM model.

\begin{deluxetable*}{ccccccc}[t!]
\setlength{\tabcolsep}{18pt}
\tablecaption{\label{tab:stats} The values of the model comparison criteria considered in this work. Here, $\Delta X$ means the difference of a comparison criterion $X$ for a model compared to the $\Lambda$CDM model. HS, ST, and ACT refer to the Hu--Sawicki, Starobinsky, and ArcTanh $f(R)$ modified gravity models, respectively.}
\tablehead{
\colhead{Model} & 
\colhead{ln$\mathcal{Z}$} & 
\colhead{AIC} & 
\colhead{BIC} &
\colhead{$\Delta$ln$\mathcal{Z}$} & 
\colhead{$\Delta$AIC} & 
\colhead{$\Delta$BIC}
}
\startdata
\multicolumn{7}{l}{\textbf{Datasets: FRBs+PantheonPlus+DESI+CC+CMB}} \\[4pt]
$\Lambda$CDM  & $-1460.58$ & $2893.50$ & $2915.38$ & $0.00$   & $0.00$   & $0.00$ \\
$w_0w_a$CDM   & $-1463.02$ & $2890.51$ & $2923.33$ & $-2.44$  & $-2.99$  & $+7.95$ \\
HS            & $-1460.51$ & $2889.38$ & $2916.74$ & $+0.07$  & $-4.11$  & $+1.36$ \\
ST            & $-1459.83$ & $2888.31$ & $2915.66$ & $+0.75$  & $-5.19$  & $+0.28$ \\
ACT           & $-1460.03$ & $2889.25$ & $2916.60$ & $-0.55$  & $-4.25$  & $+1.22$ \\
\hline  
\multicolumn{7}{l}{\textbf{Datasets: FRBs+DESY5+DESI+CC+CMB}} \\[4pt]
$\Lambda$CDM  & $-1581.44$ & $3145.86$ & $3162.65$ & $0.00$   & $0.00$   & $0.00$ \\
$w_0w_a$CDM   & $-1578.85$ & $3132.28$ & $3160.27$ & $+2.59$  & $-13.58$ & $-2.39$ \\
HS            & $-1577.54$ & $3133.96$ & $3156.35$ & $+3.90$  & $-11.90$ & $-6.30$ \\
ST            & $-1575.93$ & $3131.34$ & $3153.73$ & $+5.51$  & $-14.52$ & $-8.93$ \\
ACT           & $-1577.62$ & $3133.69$ & $3156.08$ & $+3.82$  & $-12.17$ & $-6.57$ \\
\hline
\multicolumn{7}{l}{\textbf{Datasets: FRBs+PantheonPlus+BOSS+CC+CMB}} \\[4pt]
$\Lambda$CDM  & $-1455.12$ & $2885.22$ & $2907.08$ & $0.00$   & $0.00$   & $0.00$ \\
$w_0w_a$CDM   & $-1458.36$ & $2884.82$ & $2917.62$ & $-3.24$  & $-0.40$  & $+10.54$ \\
HS            & $-1455.50$ & $2882.58$ & $2909.91$ & $-0.39$  & $-2.64$  & $+2.83$ \\
ST            & $-1455.65$ & $2882.71$ & $2910.04$ & $-0.53$  & $-2.51$  & $+2.96$ \\
ACT           & $-1456.11$ & $2882.59$ & $2909.92$ & $-0.99$  & $-2.63$  & $+2.84$ \\
\hline
\multicolumn{7}{l}{\textbf{Datasets: FRBs+DESY5+BOSS+CC+CMB}} \\[4pt]
$\Lambda$CDM  & $-1575.41$ & $3129.41$ & $3146.19$ & $0.00$   & $0.00$   & $0.00$ \\
$w_0w_a$CDM   & $-1574.86$ & $3127.52$ & $3155.50$ & $+0.55$  & $-1.89$  & $+9.30$ \\
HS            & $-1573.04$ & $3126.48$ & $3148.86$ & $+2.37$  & $-2.93$  & $+2.66$ \\
ST            & $-1572.78$ & $3126.44$ & $3148.82$ & $+2.63$  & $-2.97$  & $+2.62$ \\
ACT           & $-1573.04$ & $3126.46$ & $3148.83$ & $+2.37$  & $-2.95$  & $+2.64$ \\
\enddata
\end{deluxetable*}
\vspace{-0.8cm}

We also notice that, $\Omega_b$ and $H_0$ are well constrained in our analysis as expected, exhibiting remarkable agreement with the CMB measurements \citep{2020A&A...641A...6P} and Big Bang Nucleosynthesis (BBN) predictions \citep{2016RvMP...88a5004C}. 
By performing the test and comparing with the results obtained without including FRB data, as summarized in \autoref{tab:results} of \autoref{app:nofrb}, we find that the incorporation of FRBs leads to a significant improvement in the constraints on $H_0$ and $\Omega_b$, and a modest improvement on $\Omega_m$. Specifically, among the four data combinations considered, the inclusion of FRBs improves the constraint on $\Omega_b$ by approximately $50\%$ and on $H_0$ by about $30\sim40\%$, while the improvement for $\Omega_m$ is at the level of roughly $5\sim20\%$.
In addition, the constraints on the dark energy parameters $w_0$ and $w_a$ in dynamical dark energy models, as well as on the parameter $b$ in the $f(R)$ modified gravity model, are also moderately improved, with enhancements of about $1\sim6\%$, $5\sim13\%$, and $1\sim5\%$, respectively.
This indicates the capability of FRBs to serve as a powerful cosmological probe for the baryons in the Universe \citep{2020Natur.581..391M}, and can effectively break the parameter degeneracies when combining with other datasets.

\subsection{Model comparison}

To compare the $w_0w_a$CDM, $f(R)$, and $\Lambda$CDM models, we employ several statistical methods or criteria, including the Akaike Information Criterion (AIC) \citep{1100705}, the Bayesian Information Criterion (BIC) \citep{1978AnSta...6..461S}, and the natural logarithm of the Bayesian evidence ($\ln \mathcal{Z}$) \citep{Kass01061995}. These quantities are defined as
$
\mathrm{AIC} = -2\ln \mathcal{L} + 2k, \quad
\mathrm{BIC} = -2\ln \mathcal{L} + k\ln N, \quad
\mathcal{Z} = \int \mathcal{L}(D|\theta, M) P(\theta|M)\, d\theta,
$
where $\mathcal{L}$ is the likelihood function, $k$ is the number of free parameters, $N$ is the number of data points, and $P(\theta \mid M)$ represents the prior probability of the parameter set $\theta$ under a model $M$.

The Bayesian evidence, $\ln \mathcal{Z}$, is computed using the \texttt{dynesty} \footnote{\url{https://github.com/joshspeagle/dynesty?tab=readme-ov-file}}\citep{2020MNRAS.493.3132S,sergey_koposov_2025_17268284} package.
A larger value of $\ln \mathcal{Z}$ indicates a better model performance. For comparison of different models, we adopt a revised Jeffreys' scale \citep{Jeffreys:1939xee, Kass01061995}, according to which the strength of evidence is classified as follows: inconclusive for $0 \le |\Delta \ln \mathcal{Z}| < 1$, weak for $1 \le |\Delta \ln \mathcal{Z}| < 2.5$, moderate for $2.5 \le |\Delta \ln \mathcal{Z}| < 5$, strong for $5 \le |\Delta \ln \mathcal{Z}| < 10$, and very strong for $|\Delta \ln \mathcal{Z}| \ge 10$.

In this work, we calculate the model selection criteria discussed above for the $w_0w_a$CDM, $f(R)$, and $\Lambda$CDM models, and compare the $w_0w_a$CDM and $f(R)$ models to the $\Lambda$CDM model by computing the differences of the criteria. In Table~\ref{tab:stats}, we show the comparison results for these models.

For the $w_0w_a$CDM model, based on Bayesian evidence, we find that the dataset including PantheonPlus + BOSS provides moderate evidence in favor of $\Lambda$CDM with $\Delta \text{ln}\mathcal{Z}\simeq-3$, and the one including PantheonPlus + DESI weakly favors $\Lambda$CDM with $\Delta \text{ln}\mathcal{Z}\simeq-2$.. In contrast, for DESY5-based combinations, i.e. FRBs + DESY5 + DESI or BOSS + CC + CMB, the Bayesian evidence favors the $w_0w_a$CDM model, indicating a preference for dynamical dark energy. In this case, the dataset including DESY5 + DESI provides moderate evidence for $w_0w_a$CDM with $\Delta \text{ln}\mathcal{Z}\simeq2    .5$, whereas the one including DESY5 + BOSS inconclusively favors $w_0w_a$CDM with $\Delta \text{ln}\mathcal{Z}\simeq1$.

These results indicate that choosing  different SN~Ia and BAO datasets can affect the model selection results. In particular, combinations involving PantheonPlus and BOSS tend to favor the cosmological constant, whereas those based on DESY5 and DESI show a preference for dynamical dark energy.
Our results are in good agreement with previous similar works \citep[e.g.][]{2025arXiv250317342O,2025arXiv251008339C,2025arXiv251210585C,2025MNRAS.541.2585V,2025SCPMA..6800413H}, which are different from the results given by \citet{2025arXiv250314738D} where the Deviance Information Criterion (DIC) was employed.

For the other selection criteria shown in Table~\ref{tab:stats}, i.e. AIC and BIC, some comparison results differ from those of Bayesian evidence, since they are basically approximate statistical tool for comparison, which may introduce biases and only can be used as simple references.

For the $f(R)$ modified gravity models, compared to the $\Lambda$CDM model using Bayesian evidence, we find that when considering PantheonPlus-based combinations, i.e. FRBs + PantheonPlus + DESI or BOSS + CC + CMB, the three modified gravity models show no significant distinction from the $\Lambda$CDM model, with the Bayesian evidence remaining inconclusive ($|\Delta \ln \mathcal{Z}|=0\sim1$). However, when replacing the PantheonPlus dataset with DESY5, i.e. FRBs + DESY5 + DESI or BOSS + CC + CMB, $|\Delta \ln \mathcal{Z}|$ significantly increases. In particular, for the DESY5+DESI-based combinations, the Bayesian evidence favors the modified gravity models, with the Starobinsky model receiving strong support ($\Delta \text{ln}\mathcal{Z}\simeq5.5$) and the Hu-Sawicki and ArcTanh models having moderate evidence ($\Delta \text{ln}\mathcal{Z}\simeq4$). As for the DESY5+BOSS-based combinations, the Bayesian evidence provides weak support for the three $f(R)$ models with $\Delta \text{ln}\mathcal{Z}=2\sim3$. 

Comparing to the $w_0w_a$CDM model, we notice that, for all datasets we consider, the Bayesian evidence consistently indicates that the $f(R)$ modified gravity models are more favored, which may be attributed to the smaller number of free parameters in the $f(R)$ models compared to $w_0w_a$CDM.

\section{Summary and Conclusion}\label{sec:conclusion}
In this work, we employ different datasets of cosmological distance measurements, consisting of 112 high-quality localized FRBs, two distinct SNe~Ia compilations (PantheonPlus and DESY5), two BAO datasets (DESI-DR2 and BOSS-DR12), 39 cosmic chronometer measurements, and the angular scale of the first acoustic peak of the CMB temperature anisotropy spectrum, to constrain and compare the dynamical dark energy $w_{0}w_{a}$CDM model as well as three representative $f(R)$ modified gravity models.

By combining the FRBs, PantheonPlus, DESI, CC, and CMB datasets, the degeneracies between cosmological parameters can be effectively broken, and accurate constraint result can be obtained. We obtain tight constraints on the $w_{0}w_{a}$CDM model, yielding $w_{0} = -0.867 \pm 0.060$ and $w_{a} = -0.35^{+0.29}_{-0.25}$, which are consistent with the  results using similar datasets \citep[e.g.][]{2025ApJ...988..243L,2025PhRvD.111f1306W,2024JCAP...10..035G}. For the $f(R)$ modified gravity models, the deviation parameter $b$ is constrained to be {$b = 0.206 \pm 0.084$, $0.700^{+0.190}_{-0.130}$, and $0.198 \pm 0.081$} for the Hu-Sawicki, Starobinsky, and ArcTanh parameterizations, respectively, which are in good agreement with previous studies \citep{2023PDU....4201281K,2024MNRAS.527.7626R}. 

To further investigate the impact of different SNe~Ia and BAO datasets on the constraints, we also perform analyses with three alternative combinations, i.e. DESY5+DESI, PantheonPlus+BOSS, and DESY5+BOSS, all combined with FRBs, CC, and CMB. We find that the DESY5 and DESI datasets exhibit a stronger deviation from the $\Lambda$CDM model. In particular, for the FRBs+DESY5+DESI+CC+CMB dataset, the Bayesian evidence provides moderate or strong support for the $f(R)$ and $w_0w_a$CDM models. In contrast, the PantheonPlus and BOSS datasets relatively prefer the $\Lambda$CDM model, especially for the FRBs+PantheonPlus+BOSS+CC+CMB dataset.
We also find that the choice of the SNe~Ia dataset (PantheonPlus or DESY5) has a larger impact on the constraints than the choice of BAO dataset (BOSS or DESI). 
Besides, the Bayesian evidence indicates that all $f(R)$ modified gravity models are preferred compared to $w_{0}w_{a}$CDM for our datasets, which may be due to the fact that the $f(R)$ modified gravity models have fewer free parameters.

These findings highlight the importance of dataset selection in the joint analysis for parameter constraint and model selection, since the result can be significantly affected by the specific combination of data employed. Therefore, we should carefully choose the datasets and model selection criteria when constraining and comparing different models to obtain a reliable result.

\onecolumngrid1

\begin{acknowledgments}
S.F. and Y.G. acknowledge the support from the CAS Project for Young Scientists in Basic Research (No. YSBR-092), National Key R\&D Program of China grant Nos. 2022YFF0503404 and 2020SKA0110402. X.L.C. acknowledges the support of the National Natural Science Foundation of China through grant Nos. 11473044 and 11973047 and the Chinese Academy of Science grants ZDKYYQ20200008, QYZDJ- SSW-SLH017, XDB 23040100, and XDA15020200. This work is also supported by science research grants from the China Manned Space Project with grant Nos. CMS-CSST-2025-A02, CMS-CSST-2021-B01, and CMS-CSST-2021-A01.
\end{acknowledgments}

\appendix

\section{Constraint results without FRB dataset}\label{app:nofrb} 

\begin{deluxetable*}{ccc@{\hskip 0.8pt}c@{\hskip 0.8pt}c@{\hskip 0.8pt}cc@{\hskip 1pt}c}[ht!]
\centering

\caption{\label{tab:results} The constraint results of the cosmological parameters for different models using different datasets without FRBs. HS, ST, and ACT refer to the Hu-Sawicki, Starobinsky, and ArcTanh $f(R)$ modified gravity models, respectively.}
\tablehead{
\colhead{Model} & \colhead{$\Omega_{m}$} & \colhead{$\Omega_{b}$} & \colhead{$H_0$ (km s$^{-1}$ Mpc$^{-1}$)} & \colhead{$b$} & \colhead{$w_0$} & \colhead{$w_a$} & \colhead{$M$}
}
\startdata
\multicolumn{8}{l}{\textbf{Datasets: PantheonPlus+DESI +CC+CMB}} \\ [4pt]
$\Lambda$CDM  & $0.299\pm+0.006$ & $0.0454\pm0.0019$ & $67.32\pm1.08$ & $-$ & $-$ & $-$ & $-19.45\pm0.03$ \\
$w_0w_a$CDM   & $0.309\pm0.008$ & $0.0480\pm0.0029$ & $66.52\pm1.14$ & $-$ & $-0.862\pm0.062$ & $-0.38^{+0.33}_{-0.29}$ & $-19.45\pm0.03$ \\
HS            & $0.291^{+0.006}_{-0.007}$ & $0.0471\pm0.0019$ & $68.56\pm1.12$ & $0.218\pm0.088$ & $-$ & $-$ & $-19.45\pm0.03$ \\
ST            & $0.303^{+0.006}_{-0.008}$ & $0.0479\pm0.0022$ & $67.17\pm1.13$ & $0.700^{+0.200}_{-0.130}$ & $-$ & $-$ & $-19.45\pm0.03$ \\
ACT           & $0.292^{+0.006}_{-0.007}$ & $0.0472\pm0.0019$ & $68.44\pm1.12$ & $0.213\pm0.085$ & $-$ & $-$ & $-19.45\pm0.03$ \\
\hline
\multicolumn{8}{l}{\textbf{Datasets: DESY5+DESI+CC+CMB}} \\[4pt] 
$\Lambda$CDM  & $0.303\pm0.006$ & $0.0447^{+0.0018}_{-0.0020}$ & $66.76\pm1.10$ & $-$ & $-$ & $-$ & $-$ \\
$w_0w_a$CDM   & $0.320\pm0.008$ & $0.0481\pm0.0029$ & $65.52\pm1.14$ & $-$ & $-0.766\pm0.068$ & $-0.74^{+0.37}_{-0.32}$ & $-$ \\
HS            & $0.291^{+0.006}_{-0.007}$ & $0.0474\pm0.0019$ & $68.55\pm1.12$ & $0.306\pm0.081$ & $-$ & $-$ & $-$ \\
ST            & $0.311^{+0.007}_{-0.010}$ & $0.0490\pm0.0022$ & $66.30^{+1.30}_{-1.20}$ & $0.910\pm0.143$ & $-$ & $-$ & $-$ \\
ACT           & $0.292^{+0.006}_{-0.007}$ & $0.0475\pm0.0019$ & $68.48\pm1.12$ & $0.299\pm0.079$ & $-$ & $-$ & $-$ \\
\hline
\multicolumn{8}{l}{\textbf{Datasets: PantheonPlus+BOSS+CC+CMB}} \\[4pt]
$\Lambda$CDM  & $0.308^{+0.009}_{-0.010}$ & $0.0460\pm-0.0027$ & $66.77\pm1.40$ & $-$ & $-$ & $-$ & $-19.46\pm0.04$ \\
$w_0w_a$CDM   & $0.303^{+0.013}_{-0.012}$ & $0.0496^{+0.0033}_{-0.0046}$ & $66.23\pm1.54$ & $-$ & $-0.903^{+0.064}_{-0.080}$ & $-0.07^{+0.61}_{-0.42}$ & $-19.46\pm0.04$ \\
HS            & $0.293\pm0.011$ & $0.0462\pm0.0027$ & $68.01\pm1.51$ & $0.220\pm0.103$ & $-$ & $-$ & $-19.47\pm0.04$ \\
ST            & $0.305^{+0.009}_{-0.010}$ & $0.0472\pm0.0028$ & $66.91\pm1.45$ & $0.610^{+0.250}_{-0.140}$ & $-$ & $-$ & $-19.46\pm0.04$ \\
ACT           & $0.294\pm0.011$ & $0.0463\pm0.0027$ & $67.91\pm1.47$ & $0.210^{+0.100}_{-0.097}$ & $-$ & $-$ & $-19.46\pm0.04$ \\
\hline
\multicolumn{8}{l}{\textbf{Datasets: DESY5+BOSS+CC+CMB}} \\[4pt]
$\Lambda$CDM  & $0.316^{+0.009}_{-0.010}$ & $0.0454\pm0.027$ & $65.80\pm1.41$ & $-$ & $-$ & $-$ & $-$ \\
$w_0w_a$CDM   & $0.319^{+0.014}_{-0.012}$ & $0.0487\pm0.0036$ & $65.52\pm1.54$ & $-$ & $-0.782^{+0.080}_{-0.100}$ & $-0.70^{+0.70}_{-0.56}$ & $-$ \\
HS            & $0.292\pm0.011$ & $0.0461\pm0.0027$ & $68.03\pm1.47$ & $0.323\pm0.097$ & $-$ & $-$ & $-$ \\
ST            & $0.313^{+0.009}_{-0.012}$ & $0.0482\pm0.0031$ & $66.03^{+1.50}_{-1.30}$ & $0.850^{+0.160}_{-0.140}$ & $-$ & $-$ & $-$ \\
ACT           & $0.293\pm0.011$ & $0.0462\pm0.0027$ & $67.82\pm1.44$ & $0.309\pm0.093$ & $-$ & $-$ & $-$ \\
\enddata
\end{deluxetable*}

\onecolumngrid
\section{Observational data}\label{app:data} 
In \autoref{Tab:FRBs}, we summarize the data of 125 localized FRBs, including their redshift $z$, Right Ascension (R.A.), Declination (Decl.), observed dispersion measure ($\rm DM_{obs}$), Milky Way interstellar medium dispersion measure ($\rm DM_{MW}$) estimated using the NE2001 and YMW16 models, and the types of their host galaxies.
\startlongtable
\begin{deluxetable}{ccccccccc}
\centering
\tablecaption{The 125 localized FRBs we consider in this work, and 112 of them are selected in the constraint process. \label{Tab:FRBs}}
\tablehead{\colhead{Names} & \colhead{$z$} & \colhead{$\rm DM_{obs}$} & \colhead{$\rm R.A.$} & \colhead{Decl.} & \colhead{${\rm DM_{MW}}$ (NE2001)} & \colhead{${\rm DM_{MW}}$ (YMW16)} & \colhead{Type} & \colhead{Ref.} \\
& &$(\rm pc \, cm^{-3})$ &(deg) &(deg) &$(\rm pc \, cm^{-3})$ &$(\rm pc \, cm^{-3})$ & &}
\startdata
FRB20121102A & 0.19273 & 557.0 & 82.99460 & 33.14790 & 188.4 & 287.1 & 1 &1,2 \\
FRB20171020A & 0.00867 & 114.1 & 333.75000 & -19.66670 & 36.7 & 24.7 & 3 &4\\
FRB20180301A & 0.33040 & 536.0 & 93.22680 & 4.67110 & 151.7 & 254.0 & 1 &3\\
FRB20180814A & 0.06800 & 190.9 & 65.68330 & 73.66440 & 87.6 & 107.9 & 2 &16\\
FRB20180916B & 0.03370 & 349.3 & 29.50310 & 65.71680 & 199.0 & 324.9 & 2 &5\\
FRB20180924B & 0.32120 & 361.4 & 326.10530 & -40.90000 & 40.5 & 27.6 & 3 &6,8,10,11,25\\
FRB20181030A & 0.00385 & 103.4 & 158.58380 & 73.75140 & 41.1 & 33.0 & 1 &17\\
FRB20181112A & 0.47550 & 589.3 & 327.34850 & -52.97090 & 41.7 & 29.0 & 3 &9,10,11,25\\
FRB20181220A & 0.02746 & 208.7 & 348.69820 & 48.34210 & 118.5 & 115.3 & 3 &18\\
FRB20181223C & 0.03024 & 111.6 & 180.92070 & 27.54760 & 19.9 & 19.1 & 3 &18\\
FRB20190102C & 0.29120 & 364.5 & 322.41570 & -79.47570 & 57.4 & 43.3 & 3 &10,11,25\\
FRB20190110C & 0.12244 & 221.6 & 249.31850 & 41.44340 & 37.1 & 29.9 & 2 &19\\
FRB20190303A & 0.06400 & 223.2 & 207.99580 & 48.12110 & 29.8 & 21.8 & 2 &16\\
FRB20190418A & 0.07132 & 182.8 & 65.81230 & 16.07380 & 70.2 & 85.8 & 3 &18\\
FRB20190425A & 0.03122 & 127.8 & 255.66250 & 21.57670 & 48.7 & 38.7 & 3 &18\\
FRB20190520B & 0.24180 & 1204.7 & 240.51780 & -11.28810 & 60.2 & 50.2 & 1 &6,12\\
FRB20190523A & 0.66000 & 760.8 & 207.06500 & 72.46970 & 37.2 & 29.9 & 3 &20\\
FRB20190608B & 0.11778 & 338.7 & 334.01990 & -7.89830 & 37.3 & 26.6 & 3 &6,10,11,25\\
FRB20190611B & 0.37780 & 321.4 & 320.74560 & -79.39760 & 57.8 & 43.7 & 3 &7,10,25\\
FRB20190614D & 0.60000 & 959.2 & 65.07550 & 73.70670 & 87.8 & 108.7 & 3 &21\\
FRB20190711A & 0.52200 & 593.1 & 329.41930 & -80.35800 & 56.5 & 42.6 & 1 &7,10,25\\
FRB20190714A & 0.23650 & 504.1 & 183.97970 & -13.02100 & 38.5 & 31.2 & 3 &7,25\\
FRB20191001A & 0.23400 & 506.9 & 323.35130 & -54.74780 & 44.2 & 31.1 & 3 &7,25\\
FRB20191106C & 0.10775 & 332.2 & 199.58010 & 42.99970 & 25.0 & 20.5 & 2 &19\\
FRB20191228A & 0.24320 & 297.5 & 344.43040 & -28.59410 & 32.9 & 20.1 & 3 &3,25\\
FRB20200223B & 0.06024 & 201.8 & 8.26950 & 28.83130 & 45.6 & 37.0 & 2 &19\\
FRB20200430A & 0.16080 & 380.1 & 229.70640 & 12.37630 & 27.2 & 26.1 & 3 &7,25\\
FRB20200906A & 0.36880 & 577.8 & 53.49620 & -14.08320 & 35.8 & 37.9 & 3 &3,25\\
FRB20201020E & 0.00080 & 87.82 & 149.4863 & 
68.8256 & 40.67 & 32.23 & 2 &38 \\
FRB20201123A & 0.05070 & 433.6 & 263.67000 & -50.76000 & 251.7 & 162.7 & 1 &22\\
FRB20201124A & 0.09800 & 413.5 & 77.01460 & 26.06070 & 139.9 & 196.6 & 2 &13\\
FRB20210117A & 0.21450 & 729.1 & 339.97920 & -16.15150 & 34.4 & 23.1 & 3 &6,25\\
FRB20210320C & 0.27970 & 384.8 & 204.46080 & -16.12270 & 39.3 & 30.4 & 3 &6,25\\
FRB20210405I & 0.06600 & 565.2 & 255.33960 & -49.54520 & 516.1 & 348.7 & 3 &26\\
FRB20210410D & 0.14150 & 578.8 & 326.08630 & -79.31820 & 56.2 & 42.2 & 3 &6,14\\
FRB20210603A & 0.17720 & 500.1 & 10.27410 & 21.22630 & 39.5 & 30.8 & 3 &23\\
FRB20210807D & 0.12930 & 251.9 & 299.22140 & -0.76240 & 121.2 & 93.7 & 3 &6,25\\
FRB20211127I & 0.04690 & 234.8 & 199.80820 & -18.83780 & 42.5 & 31.5 & 3 &6,15,25\\
FRB20211203C & 0.34390 & 636.2 & 204.56250 & -31.38010 & 63.7 & 48.4 & 3 &6,25\\
FRB20211212A & 0.07070 & 206.0 & 157.35090 & 1.36090 & 38.8 & 27.5 & 3 &6,25\\
FRB20220105A & 0.27850 & 583.0 & 208.80390 & 22.46650 & 22.0 & 20.6 & 3 &6,25\\
FRB20220204A & 0.40120 & 612.6 & 274.22630 & 69.72250 & 50.7 & 46.0 & 3 &24,28,29\\
FRB20220207C & 0.04304 & 262.4 & 310.19950 & 72.88230 & 76.1 & 83.3 & 3 &24,29\\
FRB20220208A & 0.35100 & 437.0 & 319.34830 & 71.54000 & 90.4 & 107.9 & 3 &24,28,29\\
FRB20220222C & 0.85300 & 1071.2 & 203.90000 & -27.97000 & 56.00 & 42.30 & 3 &33\\
FRB20220224C & 0.62710 & 1140.2 & 166.67700 & -22.93990 & 52.54 & 51.46 & 3 &33\\
FRB20220307B & 0.24812 & 499.3 & 350.87450 & 72.19240 & 128.2 & 186.9 & 3 &24,29\\
FRB20220310F & 0.47796 & 462.2 & 134.72040 & 73.49080 & 46.3 & 39.5 & 3 &24,29\\
FRB20220319D & 0.01123 & 111.0 & 32.17790 & 71.03530 & 139.7 & 211.0 & 3 &24\\
FRB20220330D & 0.37140 & 468.1 & 165.72560 & 71.75350 & 39.0 & 30.6 & 3 &24,28,29\\
FRB20220418A & 0.62200 & 623.2 & 219.10560 & 70.09590 & 36.7 & 29.5 & 3 &24,29\\
FRB20220501C & 0.38100 & 449.5 & 352.37920 & -32.49070 & 30.6 & 14.0 & 3 &25,28\\
FRB20220506D & 0.30039 & 397.0 & 318.04480 & 72.82730 & 84.6 & 97.7 & 3 &24,28,29\\
FRB20220509G & 0.08940 & 269.5 & 282.67000 & 70.24380 & 55.6 & 52.1 & 3 &18,24,29\\
FRB20220529A & 0.18390 & 246.0 & 19.10420 & 20.63250 & 40.0 & 30.9 & 1 &27\\
FRB20220610A & 1.01600 & 1458.2 & 351.07320 & -33.51370 & 31.0 & 13.6 & 3 &25\\
FRB20220717A & 0.36295 & 637.3 & 293.30420 & -19.28770 & 118.3 & 83.2 & 3 &32\\
FRB20220725A & 0.19260 & 290.4 & 353.31520 & -35.99020 & 30.7 & 11.6 & 3 &25\\
FRB20220726A & 0.36190 & 686.2 & 73.94567 & 69.92910 & 89.5 & 111.4 & 3 &24,28,29\\
FRB20220825A & 0.24140 & 651.2 & 311.98150 & 72.58500 & 78.5 & 86.9 & 3 &24,29\\
FRB20220831A & 0.26200 & 1146.2 & 338.69550 & 70.53840 & 126.8 & 182.3 & 3 &29\\
FRB20220912A & 0.07710 & 219.5 & 347.27040 & 48.70710 & 125.2 & 122.2 & 2 &31\\
FRB20220914A & 0.11390 & 631.3 & 282.05680 & 73.33690 & 54.7 & 51.1 & 3 &24,29\\
FRB20220918A & 0.49100 & 656.8 & 17.59210 & -70.81130 & 40.8 & 28.9 & 3 &25\\
FRB20220920A & 0.15824 & 315.0 & 240.25710 & 70.91880 & 39.9 & 33.4 & 3 &24,29\\
FRB20221012A & 0.28467 & 441.1 & 280.79870 & 70.52420 & 54.3 & 50.5 & 3 &24,29\\
FRB20221027A & 0.54220 & 452.5 & 129.61040 & 71.73150 & 47.6 & 41.1 & 3 &24,28,29\\
FRB20221029A & 0.97500 & 1391.8 & 141.96340 & 72.45230 & 43.8 & 36.4 & 3 &24,28,29\\
FRB20221101B & 0.23950 & 491.6 & 342.21620 & 70.68120 & 131.2 & 192.4 & 3 &24,28,29\\
FRB20221106A & 0.20440 & 343.8 & 56.70480 & -25.56980 & 34.8 & 31.8 & 3 &24,25\\
FRB20221113A & 0.25050 & 411.0 & 71.41100 & 70.30740 & 91.7 & 115.4 & 3 &24,28,29\\
FRB20221116A & 0.27640 & 643.4 & 21.21020 & 72.65390 & 132.3 & 196.2 & 3 &28,29\\
FRB20221219A & 0.55300 & 706.7 & 257.62980 & 71.62680 & 44.4 & 38.6 & 3 &24,28,29\\
FRB20230124A & 0.09390 & 590.6 & 231.91630 & 70.96810 & 38.6 & 31.8 & 3 &24,28,29\\
FRB20230125D & 0.32650 & 640.8 & 150.20500 & -31.54000 & 88.00 & 190.79 & 3 &33\\
FRB20230203A & 0.14640 & 420.1 & 151.66159 & 35.69410 & 36.3 & 22.9 & 3 &34\\
FRB20230216A & 0.53100 & 828.0 & 155.97170 & 1.46780 & 39.5 & 28.1 & 3 &28,29\\
FRB20230222B & 0.11000 & 187.8 & 238.73912 & 30.89870 & 27.7 & 26.3 & 3 &34\\
FRB20230222A & 0.12230 & 706.1 & 106.96036 & 11.22452 & 134.2 & 188.1 & 3 &34\\
FRB20230307A & 0.27060 & 608.9 & 177.78130 & 71.69560 & 37.6 & 29.5 & 3 &28,29 \\
FRB20230311A & 0.19180 & 364.3 & 91.10966 & 55.94595 & 92.5 & 115.7 & 3 &24,28,29\\
FRB20230501A & 0.30150 & 532.5 & 340.02720 & 70.92220 & 125.7 & 180.2 & 3 &24,29\\
FRB20230506C & 0.38960 & 772.0 & 12.09980 & 42.00610 & 69.8 & 64.1 & 2 &35\\
FRB20230521B & 1.35400 & 1342.9 & 351.03600 & 71.13800 & 138.8 & 209.7 & 3 &25,29\\
FRB20230526A & 0.15700 & 361.4 & 22.23260 & -52.71730 & 31.9 & 21.9 & 3 &25\\
FRB20230613A & 0.39230 & 483.51 & 356.85000 & -27.05000 & 30.00 & 17.23 & 3 &33\\
FRB20230626A & 0.32700 & 452.7 & 235.62960 & 71.13350 & 39.3 & 32.5 & 3 &24,28,29\\
FRB20230628A & 0.12700 & 345.0 & 166.78670 & 72.28180 & 39.0 & 30.8 & 3 &24,28,29\\
FRB20230703A & 0.11840 & 291.3 & 184.62445 & 48.72993 & 26.9 & 20.7 & 3 &34\\
FRB20230708A & 0.10500 & 411.5 & 303.11550 & -55.35630 & 60.3 & 44.0 & 3 &25\\
FRB20230712A & 0.45250 & 587.6 & 167.35850 & 72.55780 & 39.2 & 30.9 & 3 &24,28,29\\
FRB20230718A & 0.03570 & 477.0 & 128.16190 & -40.45190 & 420.6 & 450.0 & 3 &25\\
FRB20230730A & 0.21150 & 312.5 & 54.66456 & 33.15930 & 85.2 & 97.4 & 3 &34\\
FRB20230808F & 0.34720 & 653.2 & 53.30405 & -51.93527 & 35.3 & 26.53 & 3 &35\\
FRB20230814B & 0.55300 & 696.4 & 335.97480 & 73.02590 & 104.8 & 137.8 & 3 &29\\
FRB20230902A & 0.36190 & 440.1 & 52.13980 & -47.33350 & 34.1 & 25.5 & 3 &25\\
FRB20230907D & 0.46380 & 1030.79 & 187.14250 & 8.65810 & 29.00 & 28.68 & 3 &33\\
FRB20230926A & 0.05530 & 222.8 & 269.12417 & 41.81000 & 52.6 & 43.7 & 3 &34\\
FRB20230930A & 0.09250 & 456.0 & 10.50000 & 41.40000 & 68.1 & 61.7 & 3 &35\\
FRB20231005A & 0.07130 & 189.4 & 246.02758 & 35.44699 & 33.5 & 28.8 & 3 &34\\
FRB20231011A & 0.07830 & 186.3 & 18.24110 & 41.74910 & 70.4 & 65.7 & 3 &34\\
FRB20231017A & 0.24500 & 344.2 & 346.75429 & 36.65268 & 64.6 & 55.6 & 3 &34\\
FRB20231020B & 0.47750 & 952.2 & 57.27830 & -37.76990 & 34.46 & 27.57 & 3 &33\\
FRB20231025B & 0.32380 & 368.7 & 270.78807 & 63.98908 & 48.6 & 43.4 & 3 &34\\
FRB20231120A & 0.03680 & 437.7 & 143.98400 & 73.28470 & 43.8 & 36.2 & 3 &24,28,29\\
FRB20231123A & 0.07290 & 302.1 & 82.62325 & 4.47554 & 89.7 & 136.9 & 3 &34\\
FRB20231123B & 0.26210 & 396.9 & 242.53820 & 70.78510 & 40.3 & 33.8 & 3 &24,28,29\\
FRB20231128A & 0.10790 & 331.6 & 199.57820 & 42.99271 & 25.0 & 20.5 & 1 &34\\
FRB20231204A & 0.06440 & 221.0 & 207.99903 & 48.11600 & 29.8 & 21.8 & 2 &34\\
FRB20231206A & 0.06590 & 457.7 & 112.44284 & 56.25627 & 59.1 & 59.3 & 3 &34\\
FRB20231220A & 0.33550 & 491.2 & 123.90870 & 73.65990 & 49.9 & 44.5 & 3 &29\\
FRB20231223C & 0.10590 & 165.8 & 259.54417 & 29.49585 & 47.9 & 38.6 & 3 &34\\
FRB20231226A & 0.15690 & 329.9 & 155.36380 & 6.11030 & 38.1 & 26.7 & 3 &25\\
FRB20231229A & 0.01900 & 198.5 & 26.46783 & 35.11292 & 58.2 & 51.8 & 3 &34\\
FRB20231230A & 0.02980 & 131.4 & 72.79761 & 2.39398 & 61.6 & 83.3 & 3 &34\\
FRB20240114A & 0.13000 & 527.6 & 321.91610 & 4.32920 & 49.7 & 38.8 & 1 &30\\
FRB20240119A & 0.37600 & 483.1 & 224.46720 & 71.61180 & 38.0 & 31.0 & 3 &29\\
FRB20240123A & 0.96800 & 1462.0 & 68.26250 & 71.94530 & 90.2 & 113.0 & 3 &29\\
FRB20240201A & 0.04273 & 374.5 & 149.90560 & 14.08800 & 38.6 & 29.1 & 3 &25\\
FRB20240209A & 0.13840 & 176.518 & 289.85000 & 86.06000 & 55.5 & 52.91 & 2 &37\\
FRB20240210A & 0.02369 & 283.7 & 8.77960 & -28.27080 & 28.7 & 17.9 & 3 &25\\
FRB20240213A & 0.11850 & 357.4 & 166.16830 & 74.07540 & 40.0 & 32.1 & 3 &29\\
FRB20240215A & 0.21000 & 549.5 & 268.44130 & 70.23240 & 47.9 & 42.8 & 3  &29\\
FRB20240229A & 0.28700 & 491.1 & 169.98350 & 70.67620 & 38.0 & 29.5 & 3 &29\\
FRB20240310A & 0.12700 & 601.8 & 17.62190 & -44.43940 & 30.1 & 19.8 & 3 &25\\
\enddata
\tablecomments{
Host galaxy types:  
(1) FRB20121102A-like repeating bursts hosted by dwarf galaxies, with stellar masses $M\sim 1-50 \times 10^7 \, M_{\odot}$ and SFRs $\sim 0.1-0.7 \, M_{\odot} \, \rm yr^{-1}$; 
(2) FRB20180916B-like repeating bursts hosted by spiral galaxies, with stellar masses $M\sim 0.1-10 \times 10^{10} \, M_{\odot}$ and SFRs $\sim 0.01-10 \, M_{\odot} \, \rm yr^{-1}$; (3) non-repeating  bursts;. 
\\
Excluded sample: The FRB20171020A, FRB20181030A, FRB20181220A, FRB20181223C,
FRB20190425A, FRB20201020E, FRB20210405I, FRB20220319D, FRB20220912A, FRB20230718A, and FRB20231230A are excluded because they do not satisfy the criterion
$\mathrm{DM}_{\mathrm{obs}} - \mathrm{DM}_{\mathrm{MW}} > 100~\mathrm{pc\,cm^{-3}}$. In addition, the FRB20190520B and FRB20220831A are removed
because they reside in extreme environments that exhibit anomalously high dispersion measures. 
\\
References: 1, \citet{2017Natur.541...58C}, 2, \citet{2017ApJ...834L...7T}, 3, \citet{2022AJ....163...69B}, 4, \citet{2018ApJ...867L..10M}, 5, \citet{2020Natur.577..190M}, 6, \citet{2023ApJ...954...80G}, 7, \citet{2020ApJ...903..152H}, 8, 
\citet{2019Sci...365..565B}, 9, \citet{2019Sci...366..231P}, 10, \citet{2020Natur.581..391M}, 11, \citet{2020ApJ...895L..37B}, 12, \citet{2022Natur.606..873N}, 13, \citet{2021ApJ...919L..23F}, 14, \citet{2023MNRAS.524.2064C}, 15, \citet{2023ApJ...949...25G}, 16, \citet{2023ApJ...950..134M}, 17, \citet{2021ApJ...919L..24B}, 18, 
\citet{2024ApJ...971L..51B}, 19, \citet{2024ApJ...961...99I}, 20, \citet{2019Natur.572..352R}, 21, \citet{2020ApJ...899..161L}, 22, \citet{2022MNRAS.514.1961R}, 23, \citet{2023arXiv230709502C}, 24, \citet{2024ApJ...964..131S}, 25, \citet{2024arXiv240802083S}, 26, \citet{2024MNRAS.527.3659D}, 27, \citet{2024arXiv241003994G}, 28, 
\citet{2024Natur.635...61S}, 29, \citet{2024arXiv240916952C}, 30, \citet{2024MNRAS.533.3174T}, 31, \citet{2023ApJ...949L...3R}, 32, \citet{2024MNRAS.532.3881R}, 33, 
\citet{2025arXiv250705982P}, 34, 
\citet{2025ApJS..280....6C}
35, 
\citet{2025arXiv250302947A}
36, 
\citet{2025MNRAS.538.1800H}
37, 
\citet{2025ApJ...979L..22E}
38,
\citet{2022Natur.602..585K}
}
\end{deluxetable}
% \vspace{-0.9cm}
In \autoref{tab:DESI}, we present BAO measurements from DESI-DR2, including the effective redshift $z_\mathrm{eff}$, the ratios $D_M/r_d$, $D_H/r_d$, $D_V/r_d$, as well as the correlation coefficient $r_{M,H}$ between $D_M/r_d$ and $D_V/r_d$.
\autoref{tab:BOSS} presents BAO measurements from BOSS-DR12, including the effective redshift $z_\mathrm{eff}$, the ratios \( D_{\rm M} \, r_{\rm d}^{\rm fid}/r_{\rm d} \), \( D_{\rm H} \, r_{\rm d}^{\rm fid}/r_{\rm d} \), where $r_{\rm d}^{\rm fid}=147.78~\mathrm{Mpc}$ is the fiducial sound horizon at the drag epoch.
Meanwhile, we list 39 measurements from cosmic chronometers, including the redshift $z$, the Hubble parameter $H(z)$, and its 1$\sigma$ uncertainty in \autoref{tab:cc}. 
\vspace{0.1cm}
\twocolumngrid

\begin{deluxetable}{cccc}
\tablecaption{DESI-DR2 BAO data used in this work.\label{tab:DESI}}
\tablehead{
\colhead{\( z_{\rm eff} \)} & \colhead{\( D_{\rm M}/r_{\rm d} \)} & \colhead{\( D_{\rm H}/r_{\rm d} \)} & \colhead{\( D_{\rm V}/r_{\rm d} \) or \( r_{\rm M,H} \)}
}
\startdata
$0.295$ & ---              & ---              & $7.942 \pm 0.075$ \\
$0.510$ & $13.588 \pm 0.167$ & $21.863 \pm 0.425$ & $-0.459$ \\
$0.706$ & $17.351 \pm 0.177$ & $19.455 \pm 0.330$ & $-0.404$ \\
$0.934$ & $21.576 \pm 0.152$ & $17.641 \pm 0.193$ & $-0.416$ \\
$1.321$ & $27.601 \pm 0.318$ & $14.176 \pm 0.221$ & $-0.434$ \\
$1.484$ & $30.512 \pm 0.760$ & $12.817 \pm 0.516$ & $-0.500$ \\
$2.330$ & $38.988 \pm 0.531$ & $8.632 \pm 0.101$ & $-0.431$ \\
\enddata
\end{deluxetable}
\vspace{-0.30cm}
\begin{deluxetable}{ccc}
\setlength{\tabcolsep}{22pt}
\tablecaption{BOSS-DR12 data used in this work. The covariance matrice can be found under the \emph{DR12 Combined-Sample Consensus Results} section at \url{https://www.sdss3.org/science/boss_publications.php}.\label{tab:BOSS}}
\tablehead{
\colhead{$z_\mathrm{eff}$} & \colhead{\( D_{\rm M} \, r_{\rm d}^{\rm fid}/r_{\rm d} \)} & \colhead{\( D_{\rm H} \, r_{\rm d}^{\rm fid}/r_{\rm d} \)}
}
\startdata
0.38 & 1512.39 & 81.2087 \\
0.51 & 1975.22 & 90.9029 \\
0.61 & 2306.68 & 98.9647 \\
\enddata
\end{deluxetable}

\begin{deluxetable}{ccc}
\tablecaption{The Cosmic Chronometer data used in this work.\label{tab:cc}}
\tablehead{
\colhead{$z$} & \colhead{$H(z)$} & \colhead{Ref.}
}
\startdata
0.09    & 69 $\pm$ 12      & \cite{Stern:2009ep} \\
0.17    & 83 $\pm$ 8       & \cite{Stern:2009ep} \\
0.27    & 77 $\pm$ 14      & \cite{Stern:2009ep} \\
0.40    & 95 $\pm$ 17      & \cite{Stern:2009ep} \\
0.48    & 97 $\pm$ 62      & \cite{Stern:2009ep} \\
0.88    & 90 $\pm$ 40      & \cite{Stern:2009ep} \\
0.90    & 117 $\pm$ 23     & \cite{Stern:2009ep} \\
1.30    & 168 $\pm$ 17     & \cite{Stern:2009ep} \\
1.43    & 177 $\pm$ 18     & \cite{Stern:2009ep} \\
1.53    & 140 $\pm$ 14     & \cite{Stern:2009ep} \\
1.75    & 202 $\pm$ 40     & \cite{Stern:2009ep} \\
0.44    & 82.6 $\pm$ 7.8   & \cite{Blake:2012pj} \\
0.60    & 87.9 $\pm$ 6.1   & \cite{Blake:2012pj} \\
0.73    & 97.3 $\pm$ 7.0   & \cite{Blake:2012pj} \\
0.179   & 75 $\pm$ 4       & \cite{Moresco:2012jh} \\
0.199   & 75.0 $\pm$ 5     & \cite{Moresco:2012jh} \\
0.352   & 83.0 $\pm$ 14    & \cite{Moresco:2012jh} \\
0.593   & 104.0 $\pm$ 13   & \cite{Moresco:2012jh} \\
0.68    & 92.0 $\pm$ 8     & \cite{Moresco:2012jh} \\
0.781   & 105.0 $\pm$ 12   & \cite{Moresco:2012jh} \\
0.875   & 125.0 $\pm$ 17    & \cite{Moresco:2012jh} \\
1.037   & 154.0 $\pm$ 20    & \cite{Moresco:2012jh} \\
0.35    & 82.7 $\pm$ 8.4    & \cite{Chuang:2012qt} \\
0.07    & 69.0 $\pm$ 19.6   & \cite{Zhang:2012mp} \\
0.12    & 68.6 $\pm$ 26.2   & \cite{Zhang:2012mp} \\
0.20    & 72.9 $\pm$ 29.6   & \cite{Zhang:2012mp} \\
0.28    & 88.8 $\pm$ 36.6   & \cite{Zhang:2012mp} \\
0.57    & 96.8 $\pm$ 3.4    & \cite{BOSS:2013rlg} \\
2.34    & 222.0 $\pm$ 7.0   & \cite{BOSS:2014hwf} \\
1.363   & 160.0 $\pm$ 33.6  & \cite{Moresco:2015cya} \\
1.965   & 186.5 $\pm$ 50.4  & \cite{Moresco:2015cya} \\
0.3802  & 83.0 $\pm$ 13.5   & \cite{Moresco:2016mzx} \\
0.4004  & 77.0 $\pm$ 10.2   & \cite{Moresco:2016mzx} \\
0.4247  & 87.1 $\pm$ 11.2   & \cite{Moresco:2016mzx} \\
0.4497  & 92.8 $\pm$ 12.9   & \cite{Moresco:2016mzx} \\
0.4783  & 80.9 $\pm$ 9.0    & \cite{Moresco:2016mzx} \\
0.47    & 89 $\pm$ 50       & \cite{Ratsimbazafy:2017vga} \\
0.75    & 98.8 $\pm$ 33.6   & \cite{Borghi:2021rft} \\
0.80    & 113.1 $\pm$ 20.73 & \cite{Jiao:2022aep} \\
\enddata
\end{deluxetable}

%\clearpage
\vspace{-2cm}
\bibliographystyle{aasjournal} 
\bibliography{main}

\end{document}